\documentclass[twocolumn]{aa}
\usepackage{graphicx}
\input epsf.sty
\newcommand{\sax}{{\it BeppoSAX}}

\begin{document}
\title{Slow and fast components in the X-ray light curves of Gamma-Ray
Bursts}
\author{L.~Vetere\inst{1,2}
\and E.~Massaro\inst{1}
\and E.~Costa\inst{2}
\and P.~Soffitta\inst{2}
\and G.~Ventura\inst{3}
}
\institute{
Dipartimento di Fisica, Universit\`a La
Sapienza, Piazzale A. Moro 2, I-00185 Roma, Italy
\and INAF, IASF - Sezione di Roma, via del Fosso del Cavaliere,
I-00133 Roma, Italy
\and Stazione Astronomica di Vallinfreda, via del Tramonto,
I-00020 Vallinfreda (RM), Italy
}
\offprints{enrico.massaro@uniroma1.it}
\date{Received ....; accepted ....}
\markboth{L. Vetere et al.: Slow and fast components in the X-ray light curves of Gamma-Ray
Bursts}
{L. Vetere et al.: Slow and fast components in the X-ray light curves of Gamma-Ray
Bursts}
\abstract{ Gamma-ray burst light curves show quite different
patterns: from very simple to extremely complex. 
We present a temporal and spectral study of the light curves in three
energy bands (2--5, 5--10, 10--26 keV)
of ten GRBs detected by the Wide Field Cameras on board \sax. 
For some events the time profiles are characterized by peaks
superposed on a slowly evolving pedestal, which in some cases
becomes less apparent at higher energies. 
We describe this behaviour with the presence of two components 
(slow and fast) having different variability time scales. 
We modelled the time evolution of slow components by means of an 
analytical function able to describe asymmetric rising and decaying
profiles. The residual light curves, after the subtraction of
the slow components, generally show structures more similar to
the original curves in the highest energy band.
Spectral study of these two components was performed 
evaluating their hardness ratios, used also to derive photon
indices.
Slow components are found generally softer than the fast ones
suggesting that their origin is likely different.
Being typical photon indices lower than those of the afterglows
there is no evidence that the emission processes are similar.
Another interesting possibility is that slow components can be
related to the presence of a hot photosphere having a thermal
spectrum with $kT$ around a few keV superposed to a rapid variable
non-thermal emission of the fast component.
}
\authorrunning{L. Vetere et al.}
\titlerunning{GRB X-ray light curves}

\maketitle


\section{Introduction}

Light curves (hereafter LC) of Gamma-ray Bursts (GRB) are
characterized by a large variety of shapes: they can be very
simple with a unique prominent peak or extremely complex with
several peaks having different durations, heights and spectra.
These particular behaviours have been extensively studied looking
for possible relations between different parameters of the LCs
(see Fenimore et al. 1995, Norris et al. 2000, Reichart et al.
2001). The majority of GRBs has been detected at photon energies
higher than about 30 keV, and only a relatively small set of data
in the energy band of a few keV is available. In particular the
two Wide Field Cameras (WFC) (Jager et al. 1997), on board the
satellite \sax~ (Boella et al. 1997), detected several events in
the energy range 2--26 keV and provided some detailed X-ray LCs of
the prompt emission. Their shapes confirm the large variability
with the energy and the different spectral properties of the
various substructures. These differences become more relevant when
comparing the shape of the X-ray curve of a burst with that
observed in the low energy $\gamma$-rays either by the \sax~
Gamma-Ray Burst Monitor (GRBM) (Frontera et al. 1997) or by {\it
BATSE} (Fishman et al. 1985).

A seminal work on BATSE LCs was done by Norris et al. (1996) who
deconvolved the temporal profiles of 46 GRBs into pulses using
a model-dependent, least squared pulse fitting algorithm in four
energy bands from 25 keV to 1 MeV. Their basic assumption was
that structured burst emission is the superposition of a series 
of individual pulses.
Using a pulse model characterised by separate rise and decay time 
constants and a sharpness peak parameter Norris et al. (1996) found 
different trends of pulse shape with energy.
They found that the widths of individual pulses have a power law 
dependence on energy consistent with the result by Fenimore et al.
(1995). 
Furthermore, they examined the pulse shape dependence on
energy and introduced the following 'pulse paradigm' between pulse 
asymmetry, width and spectral softening: 
among pulses, with relatively well defined shapes, a decreasing 
rise-to-decay ratio corresponds, for lowering energies, to wider and 
more asymmetric shapes with the pulse centroid shifted at later times. 
These results are relevant because they can lead to major understanding 
of the physics of GRBs. 
However, Norris et al. (1996) succeeded in deconvolving 41 light curves 
in different pulses while the fits were not successful for 5 GRBs. 
In fact, LCs of these GRBs, especially in the lowest energy range 
(25--55 keV), seemed to be composed by several pulses superposed on 
the top of a broader, smooth component.

A direct inspection of the LCs of GRBs detected in the WFCs shows
that in some of them the lowest level of the count rate, from
which narrow peaks are emerging, is significantly higher than the
average background level during the entire duration of the burst.
For several GRBs the amplitude of this effect decreases at higher
energies, although in a few cases it is increasing or remains
stable. This particular structure of the X-ray LCs of bright GRBs
suggests the possibility that the time evolution of the events can
be described in terms of two components: one showing the rapid
evolving peaks and spikes and the other with a much longer typical
time scale of the order of several tens of seconds. 
The occurrence of an underlying X-ray component with a spectral
distribution softer than the mean prompt emission has also been
recently pointed out by Vanderspek et al. (2004) for \object{GRB 030329}
observed by $HETE$ (Shirasaki et al. 1999).
We will show in the present paper, based on an analysis of the
\sax~ WFC observations, that this is not a unique case but
there are several GRBs with a similar behaviour. Likely, these
events constitute a GRB subclass whose main characteristics must
be investigated.

\begin{table*}
\centering
\caption{ Main parameters of GRBs observed with BeppoSAX WFCs and
considered in our analysis of slow components. X-Peak Flux are 
measured in 2--10 keV energy band and  $\gamma$-Peak Flux in 
40--700 keV.
} \label{tab1}
\begin{tabular}{lllccccc}
\hline
 GRB & ~~$RA$ & ~~$DEC$ & Duration & X-Peak Flux$^{(a)}$ & $\gamma$-Peak 
Flux$^{(a)}$ & $z$ & afterglow \\
 & h~~ m~~ s  & ~~$^{\circ}$~~~ $'$~~~ $''$ & s~ & 10$^{-7}$erg cm$^{-2}$
& 10$^{-7}$erg cm$^{-2}$ &   &  \\ \hline
\object{GRB 980519} & 23 22 15  & +77 15 00  & 250 & 0.51 & 13  &  & X-O-Radio\\
\object{GRB 990704} & 12 19 30  & $-$03 48.2 & 30  & 1.0  & 1.8 &  & X \\
\object{GRB 001011} & 18 23 4.6 & $-$50 54 16& 50  & 0.07 & 25  &  & O \\ 
\object{GRB 001109} & 18 30 02  & +55 18.4   & 100 & 0.22 & 4.2 &  & X \\
\object{GRB 010222} &14 52 12.55& +43 01 6.3 & 200 & 2.1  & 86  & 1.47& X-O\\
\object{GRB 010412} & 19 23 36  & +13 37     & 100 & 0.46 & 17  &  &$(\star)$\\ \hline
\object{GRB 990123} & 15 25 30.6& +44 46 00 &$\geq$70& 0.5 &170 & 1.6 & X-O\\
\object{GRB 990705} & 05 09 52  &$-$72 08 00 & 50  & 0.85 & 37  & 0.843 & X-O\\ 
\object{GRB 990908} & 06 53 & $-$75          & 150 & 0.17 & 1.0 &  &$(\star)$\\ 
\object{GRB 000528} & 10 45 24  & $-$34 00   & 130 & 0.21 & 14  &  & X\\ \hline
\multicolumn{8}{c} { }
\end{tabular}

$(a)$: (Frontera 2004)
$(\star)$: No follow-up performed with ~\sax ~NFI 
\end{table*}

Using the WFC database, containing data on 56 GRBs, we derived the
LCs in three energy bands and selected those showing a presence of
a possible component evolving over the entire duration of the
burst. In the following we will refer to it as Slow Component
(shortly SC), while the residual  prompt emission structure, often
characterised by a series of short duration peaks will be
indicated as Fast Component (FC),  and the original light curve will be
indicated as OR.  
This approach is not in contrast with that of Norris et al. (1996). 
In fact, they are just two different points of view: our attention 
is focused on the broad and smooth structures relevant in the 
X-ray energy range, while their work is centred on the modelling of 
the fast and spiky features in the $\gamma$-ray band.
We found that about one third of the database shows a possible SC 
while 13 GRBs have light curves too faint to tell something about.

In this paper we present the results on ten GRBs for which it is
possible to distinguish the two components. These events were
selected to have a representative sample of different types of
SCs.
LCs and other data on the remaining GRBs in the WFC database will
be described in subsequent paper.

We used a simple analytical model to describe the time evolution
of the SCs and to derive their main temporal and spectral properties.
We found that spectral properties of SC and FC of the same GRB are
not generally the same suggesting different emission processes.
We also searched for possible connections between the SCs and the
subsequent X-ray afterglow emission and discuss some possible
physical scenarios to understand the origin of this phenomenology.
In particular, we speculate that SC can be related to an outflow
photosphere, early proposed by M\'esz\'aros, Laguna \& Rees (1993)
and developed by M\'esz\'aros \& Rees (2000), M\'esz\'aros et al. (2002),
Rees \& M\'esz\'aros (2005).
The result of our spectral analysis suggest that several SCs have
softening spectra, likely associated with a thermal emission.

\section{Observations and data reduction}

All the data analysed in this work were taken with the two WFCs on
board the \sax~ satellite. Each WFC is a multi-wire proportional
counter with an open area of 25$\times$25 cm$^{2}$ and a coded
mask in the field of view allows to obtain images of the sky
(Jager et al. 1997). The transparency of the mask is 33\% and the
flux is limited by a collimator with no flat field. The effective
area is $\sim$140 cm$^{2}$ in the centre of the field of view.
Most GRBs were observed with an effective area of a few tens of
cm$^{2}$. There are two different ways to extract a LC from WFC
data:
\begin{itemize}
\item  considering the whole detector image and deconvolving it using 
the mask response function with an implicit background subtraction
\item selecting just the region of the detector concerned with the
burst, without subtracting the background component (i.e. excluding
the pixels occulted by the collimator).
\end{itemize}

The former way gives LCs without the background, while the latter
gives LCs in which counts are distributed according  to Poisson
statistics.  It is important to note that generally the
integrated count rate of instrumental  background is $\sim$130
counts/s and it is due to the cosmic X-ray background and undetected
sources for almost 85\%.
In some regions of the sky rich in X-ray sources, like the Galactic
Centre or Cygnus, this value can even double.  
Furthermore, the burst's counts depend not only on the incident
spectrum but also on the mask transparency, the detector efficiency 
and on the burst position in the field of view.
For this reason we decided to use the former way of LC
extraction to have a more controlled background subtraction,
in particular when a bright source is in the field of view.
 For the majority of the events considered in this work 
the count rate was much above the background one so that its 
subtraction was straightforward and did not affect the burst's 
light curve.

For each burst we created LCs in three energy bands: 2--5, 5--10,
10--26 keV, selected to have approximately the same number of
counts in each band for a typical GRB. The bin integration time
was generally 0.5 s, but for few bursts with a low S/N ratio a bin
time of 1 s was used. The main properties of the 10 GRBs
considered in the present paper are given in Table 1.

\subsection{Spectral analysis}
 The evaluation of spectral properties of SCs is limited by the three
energy bands used. The simplest way to obtain useful information is
to compute the Hardness Ratios (hereafter HR) between the various bands
which were used to estimate a photon index $\Gamma$ by convolving a power
law input spectrum with the WFC response matrix in the GRB direction.
Using the counts in two selected bands we evaluated the expected HR
from a set of spectra with photon indices $\Gamma$ assigned in a
given range simulated by means the {\it XSPEC} code (Arnaud et al.
1996).
In this way we obtained a relationship between $\Gamma$ and the HR,
from which we could find the photon index corresponding to the
observed HR.
We calculated the values of HR and $\Gamma$ for the original (OR) LCs
and for both the FC and SC of all GRBs.
Hereafter we will indicate with $HR_1$ the HR between 5--10 keV and
2--5 keV energy range and with $HR_2$ the HR between 10--26 keV and
5--10 keV, the corresponding photon indices are indicated as $\Gamma_1$
and $\Gamma_2$, respectively.
Statistical uncertainties were computed by propagating the 1$\sigma$
errors on HRs with the usual quadratic formulae. When working with
time integrated LC, the number of counts is high and the resulting
errors are generally smaller than those obtained by spectral best
fits over several energy bins.
Reported photon indices must therefore be considered as a quantitative
indication of the mean spectral behaviour in the WFC range.

\section{Slow component modelling}

There is no direct (and unambiguous) way to separate SC and FC in a
X-ray LC of a GRB.
We therefore applied a simple heuristic approach and modeled the
SC by means of the following analytical formula containing only
a small number of parameters:
\begin{equation}\label{model}
F_S(t,E_n)=A(E_n)~ (t-t_{0})^b~ exp[-C~(t-t_{0})^s]
\end{equation}
where $F_S(t,E_n)$ is the time dependent number of counts in the
energy band $n$ (hereafter we use $n$=1, 2, 3 for the lowest, mid
and highest energy bands specified in Sect. 2, respectively), $A(E_n)$
is the amplitude of the SC in the considered energy band, 
and the values of remaining shape parameters $b$, $C$ and $s$ were 
chosen by interpolating the local minima of the LC in the band where 
the SC was more apparent. Parameters' uncertainties cannot be 
evaluated in a simple way because they were not found by means of
a statistical regression. For the GRBs with a high S/N ratio, however,
we tried to obtain an indicative range for the parameters by fitting
our SC model to those data points for which a zero residual is expected.
Typically , we obtained good $\chi_{r}^{2}$ and the percent errors
were $\sim$ 20\%.

An advantage of Eq.~(\ref{model}) is the possibility to
represent rising and decaying sections having different time
scales with a unique smooth function. We are aware that this
description of SCs is somewhat arbitrary and that the values of
the involved parameters cannot be estimated on the basis of a
statistical method. The main fact is that there is no `a priori'
information on possible physical models of GRBs producing a SC and
on its spectral and temporal evolution. The simple approach followed
by us should be therefore considered a first tool to distinguish
phenomena characterised by different time scales in the prompt emission
LCs.

 In most cases we found that the assumption of the energy 
independence of the time evolution of the SC was substantially 
confirmed, and in only two cases an energy dependent time 
evolution was needed. We discuss these cases when
presenting the results on individual bursts. We found also that
some possible residuals of a SC are apparent only in a small
number of LCs and, when present, these residuals are
generally small indicating that a substantial fraction of SC was
actually subtracted. In the majority of cases SCs are soft and
quite small or unapparent in the highest energy band, however, for
about one third of the studied GRBs the SCs is growing from 2--5 keV to
5--10 keV and to 10--26 keV. In the next section we describe in
detail the LCs and the SCs of all the GRBs.

Residual FCs were obtained by subtracting the SCs from the original
data sets:
\begin{equation}\label{FC}
F_F(t, E_n) = F(t, E_n) - k_{ni}~ F_S(t-t_{\star}, E_i)
\end{equation}
where $F(t, E_n)$ is the original LC, $F_F$ the fast component,
and $k_{ni}=A(E_n)/A(E_i)$ is the ratio between the amplitudes for
the energy bands $n$ and $i$ (usually $i=1$), where parameters were
estimated; $t_{\star}$ is a possible time delay between the
SCs at different energies: it was taken different from zero only
for \object{GRB~000528}. A useful check `a posteriori' of our modelling was
done by comparing the FC light curves of the same GRB at different
energies resulting with very similar profiles. In only one
case, we did not obtain an acceptable model and we had to model it
using the sum of two curves like Eq.~(\ref{model}) shifted
in time.
\begin{table*}
\centering
\caption{ Values of the parameters used for the analytical law to
estimate the slow component at. $A$ is the amplitude of
the SC at 2-5 keV, and $k_{21}=A_{E{_2}}/A_{E_{1}}$,
$k_{31}=A_{E{_3}}/A_{E_{1}}$. } \label{tab2}
\begin{tabular}{cccccccc}

\hline GRB & $A$ & $b$ & $C$ & $s$ & $k_{21}$ & $k_{31}$&$t_{0}$ (s)\\
\hline
\object{GRB 980519} & 5.0~$10^{-9}$ & 5.9  & 0.10 & 1 & 0.60 &  & $-$60.84\\
\object{GRB 990704} & 2.15 & 0  & 1.2e-4 & 3.7 &  &  & $-$1.3\\
       & 1.6  & 0  & 4e-4   & 3.7 &  &  & $-$3.3 $(a)$\\
\object{GRB 001011} & 6.0~$10^{-7}$ & 4.9  & 0.12 & 1 & 0.67 &  & $-$10.8\\
\object{GRB 001109} & 4.0~$10^{-8}$ & 5.4 & 0.12 & 1 & 0.75 &  & $-$10.41\\
\object{GRB 010222} & 1.83~$10^{-4}$ & 3.45  & 0.0878 & 0.95 & 0.80 & &163.6 \\
\object{GRB 010412} & 2.5~$10^{-7}$ & 4.9 & 0.10 & 1 & 0.80 &  & 15.58\\
\hline
\object{GRB 990123} & 2.8~$10^{-5}$ & 4  & 0.112 & 1 & 1 & 1& $-$32.0\\
       & 8.8~$10^{-4}$ & 3.05 & 0.085 & 1.07 & 1.125 & 0.82 & 0\\
\object{GRB 990705} & 0.5~$10^{-9}$ & 7.9  & 0.2 & 1 & 1.4 & 1.7&$-$45.58 \\
\object{GRB 990908} & 0.7~$10^{-9}$ & 6.4  & 0.10 & 1 & 0.86 & 0.5& $-$50.0\\
\object{GRB 000528} & 8.0~$10^{-9}$ & 5.8  & 0.11 & 1 & 2 & 2 & $-$30.598\\
\hline
\multicolumn{8}{c} { }
\end{tabular}

$(a)$ -- Values for the 5--10 keV SC.
\end{table*}

It is known that GRB peaks tend to be broader at lower energies.
This effect was studied at first by Fishman et al. (1992) who
noticed that individual peaks frequently are narrower and better
defined at higher energies.
After Link, Epstein, \& Priedhorsky (1993) showed that this is a
prevalent property of most bursts and Fenimore et al. (1995),
using the average autocorrelation function and the average pulse
width, showed there is a well defined relationship between the
latter and the energy:
\begin{equation}  \label{Fenimore}
\Delta \tau \propto E^{-0.45}
\end{equation}
where $\Delta \tau$ is the FWHM of each individual peak. This
result was in agreement with the theoretical
expectations of the shock emission model (Tavani 1996)
and with the pulse modelling by Norris et al. (1996).
A possible consequence of this low energy broadening is that peaks
with a small time separation can blend together and produce an
apparent underlying pedestal evolving on a longer time scale.
As discussed in Sect. 6 we verified that the SC is not originated
by this effect.

We also analyzed the distribution of the power on the various time
scales of the SCs and FCs of a couple of GRBs having a high S/N ratio
by means of wavelet analysis. 
This method provides useful information on the power distribution of
the signals in the various time scales from those of the narrowest spikes 
to the entire duration of the prompt emission. Moreover, it can 
also be used to compare the power distributions between the original 
LCs and their components.
Standard wavelet analysis is based on the
computation of Wavelet Power Spectra defined as the normalised square
of the modulus of the wavelet transform:
\begin{equation}\label{wav1}
W_{l,m} = \zeta  \left|w_{l}(a_{m})\right|^2
\end{equation}
where
\begin{equation}\label{wav2}
w_{l}(a_{m}) = \sqrt{\frac{\Delta t}{a_{m}}}~ \Sigma_i~ x_i~
\psi^*\Big( \frac{(i-l) \Delta t}{a_{m}}\Big)
\end{equation}
with $\Delta t$ the sampling time of the series $x_i$,
$\psi^*$ the complex conjugate of the wavelet function,
$\zeta$ a normalisation factor and $a_m$ the time scale.
For a brief and practical introduction to wavelet analysis
and its computation we refer to Torrence and Compo (1998)
while for application to astrophysical data to the recent
paper by Lachowicz and Czerny (2005).

In our analysis we adopted the DOG2 wavelet (second Derivative
of a Gaussian, also known as Mexican Hat or Marr wavelet) which
gives a good description of the scale of individual pulses.
To have a more representative picture of the signal strength
distribution over the various time scale we preferred to plot
instead of $W_{l,m}$ only the positive values of
$\sqrt{\zeta}~w_{l}(a_{m})$.
In this way we excluded from the power distribution all the
time intervals where the negative wings of the DOG2 wavelet
are in anticorrelation with the signal.
In the following we will refer to these plots as Wavelet Positive
Amplitude Spectra (WPAS).

\begin{figure*}
 \begin{center}
\hspace{0.cm}
 \begin{tabular}{cc}
         \begin{minipage}{0.48\textwidth}
                 \begin{center}
                 \includegraphics[height=8cm,angle=-90]{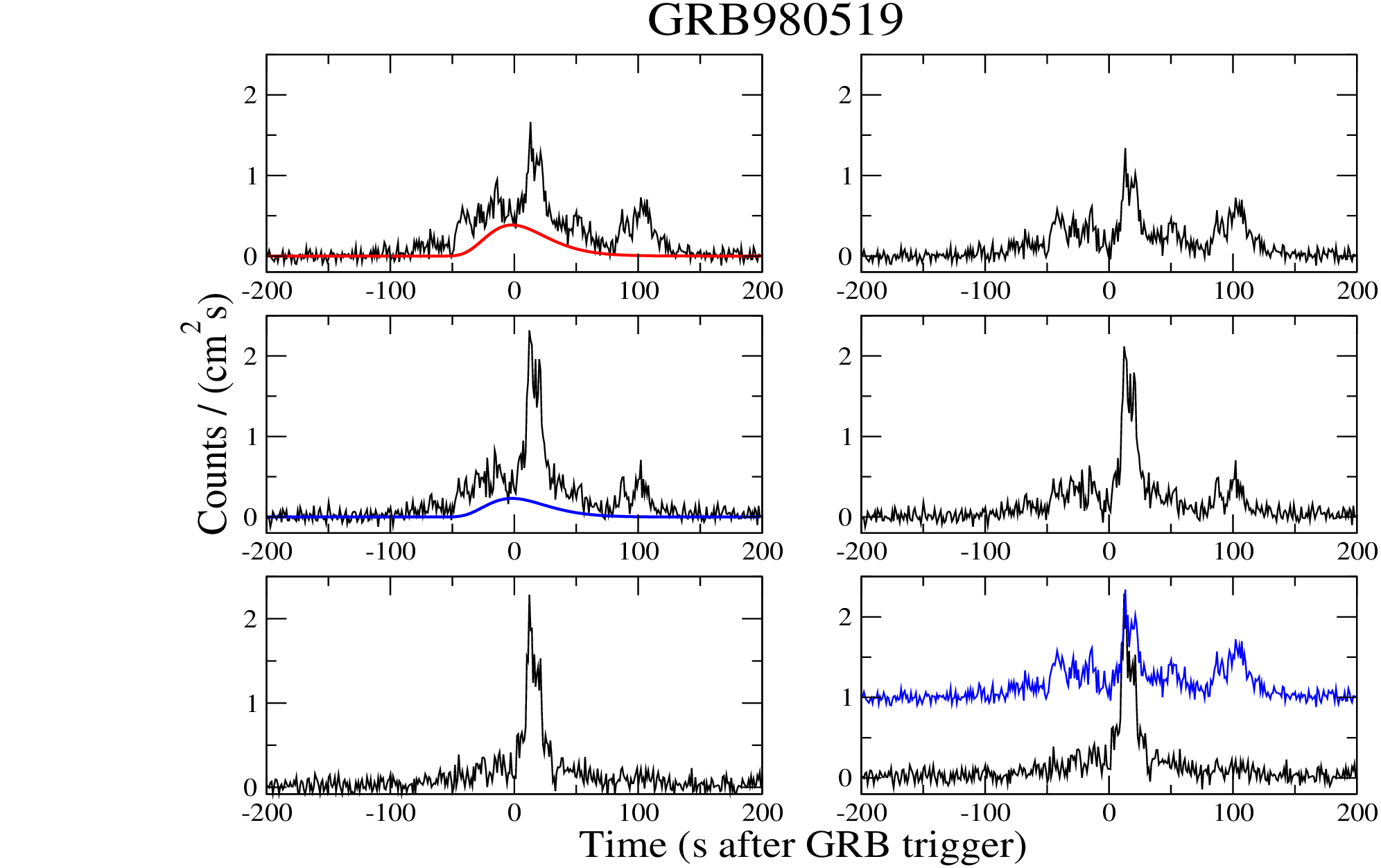}
                 \end{center}
         \end{minipage}
         &
         \begin{minipage}{0.48\textwidth}
                 \begin{center}
                 \includegraphics[height=8cm,angle=-90]{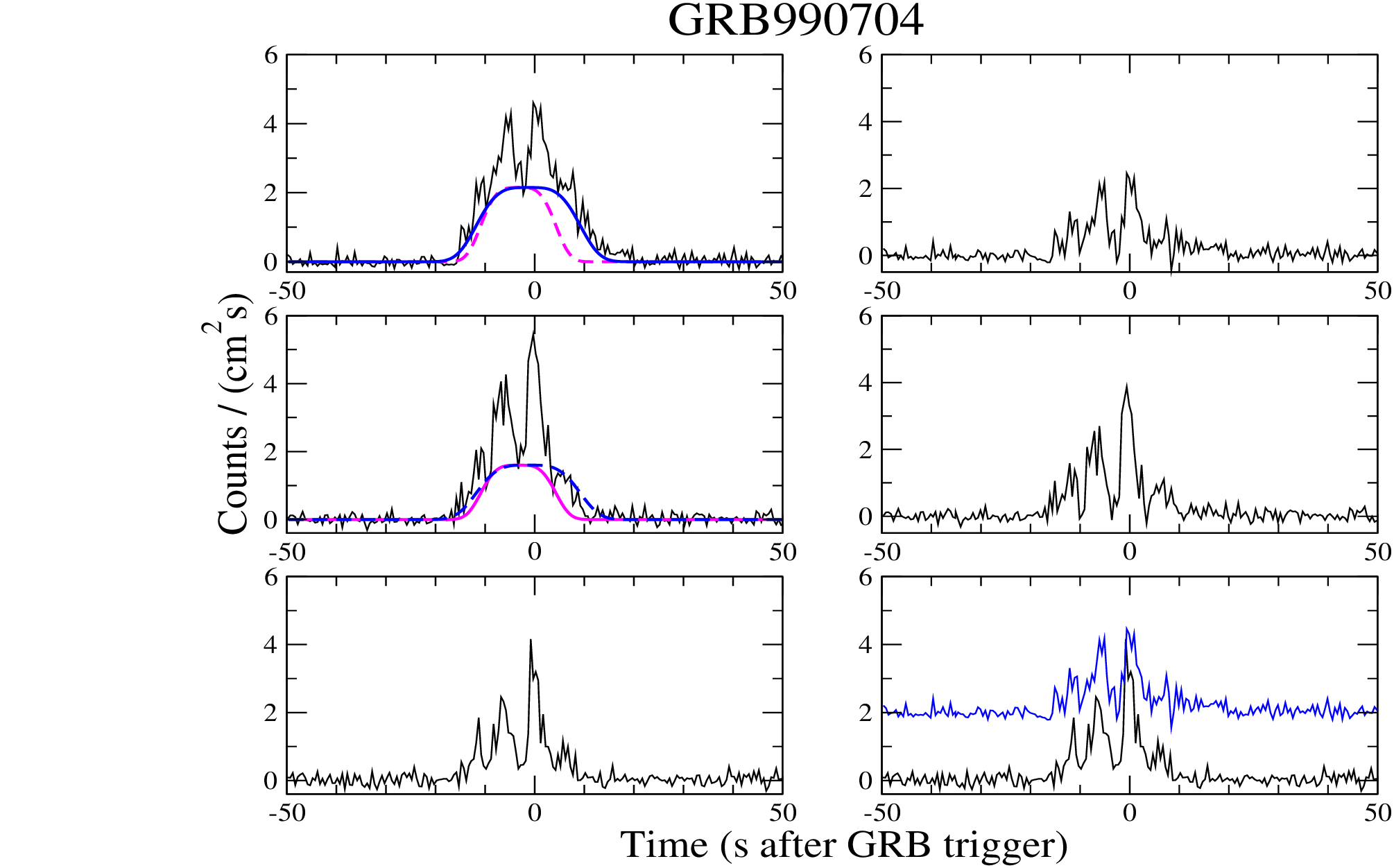}
                 \end{center}
         \end{minipage}
         \\
         \begin{minipage}{0.48\textwidth}
                \caption{\label{fig1}X-ray light curves of GRB~980519
                  in the three considered energy ranges.
                  The top and middle left  panels show the total counts
                  and the SC model; the top and middle right panels
                  show the corresponding FCs after the SC subtraction.
                  The lowest two panels show on the left the 10--26 keV LC
                  and on the right side the same curve with superposed
                  on it the FC at 2--5 keV shifted of 1.}
        \end{minipage}
       &
         \begin{minipage}{0.48\textwidth}
               \caption{\label{fig2} X-ray light curves of GRB 990704 
                  in the three considered energy ranges.
                  The top and middle left  panels show the total counts
                  and the used SC model (solid line). Dashed lines are 
		  the SC used in the other band scaled in amplitude just 
		  to show the difference of the two models; 
		  the top and middle right panels show the corresponding 
		  FCs after the SC subtraction.
                  The lowest two panels show on the left the 10--26 keV LC
                  and on the right side the same curve with superposed
                  on it the FC at 2--5 keV shifted of 1. }
         \end{minipage}
 \end{tabular}
 \end{center}
 \end{figure*}

\section{Description of individual GRB}

As written above the GRBs analysed in this work can be divided in
two groups characterised by the presence or not of a SC in the
highest energy band (10--26 keV), hereafter named {\it hard} 
and {\it soft} SC, respectively.
In the case of \object{GRB~010222}, however, we cannot exclude that a SC
is marginally detectable in the highest energy range. This burst
was one of the brightest recorded events and it is possible that
the SC appearance is due to the high S/N ratio. Although its
behaviour is intermediate between soft and hard SC GRBs we classified
it in the former type because the SC in the 10--26 keV range, if
present, must be much lower than below 10 keV.
In the next subsections we describe the analysis done
for each burst, considering first those without an apparent
10--26 keV SC.

\subsection{GRBs with soft Slow Components}

\subsubsection{\object{GRB~980519}}
\object{GRB 980519} was detected on 1998 May 19, 12:20:13 UT by CGRO-BATSE
and \sax -GRBM. It was in the field of view of the \sax -WFC 2,
allowing an estimate of its position within a 3$'$ error circle
(Piro 1998). Jaunsen et al. (1998) detected in the WFC error box a
fading optical counterpart in $V$ and $I$ bands 8.8 h after the
burst (see also Djorgovski et al. 1998, Hjorth et al. 1999a). A
fading X-ray afterglow was detected 9.7 h after the burst with the
Narrow-Field Instruments (NFI) on \sax~ (Nicastro et al. 1998).

The LC at 2--5 keV (Fig.~\ref{fig1}, upper left panel) is
characterised by a prominent peak at about the centre of the
 prompt emission and by some others of smaller amplitude before and after
the main peak. The time structure of the evolving SC is not
simple: it is detectable in the 2--5 keV data, but at 5--10 keV is
less apparent and it is undetectable at 10--26 keV. The values of
the parameters of Eq.~(\ref{model}) were therefore estimated from
the 2--5 keV LC and then we subtracted the resulting SC from the
original one to derive the FC. The 5-10 keV SC was obtained by a
simple amplitude scaling. The SC parameters are given in Table
~\ref{tab2}, and the fraction of counts is reported in Table
~\ref{tab3}.

The comparison of the LC at 10--26 keV with the FCs shows that
they really look alike indicating that the SCs modelling is
acceptable (Fig.~\ref{fig1} lower right panel). We cannot exclude,
however, that SC may be more complex and structured. Note that
above 10 keV only the central peak is clearly apparent, indicating
that it has a spectrum harder than the other features, which at
these energies are practically absent. A soft SC could not consist
of only a simple and regular profile and can be due to a couple of
features, one before and the other after the main peak. In this
case the fractions of counts in Table ~\ref{tab3} should be
properly considered as lower limits to the actual SC content.

\begin{figure*}
 \begin{center}
\hspace{0.cm}
 \begin{tabular}{cc}
         \begin{minipage}{0.48\textwidth}
                 \begin{center}
                 \includegraphics[height=8cm,angle=-90]{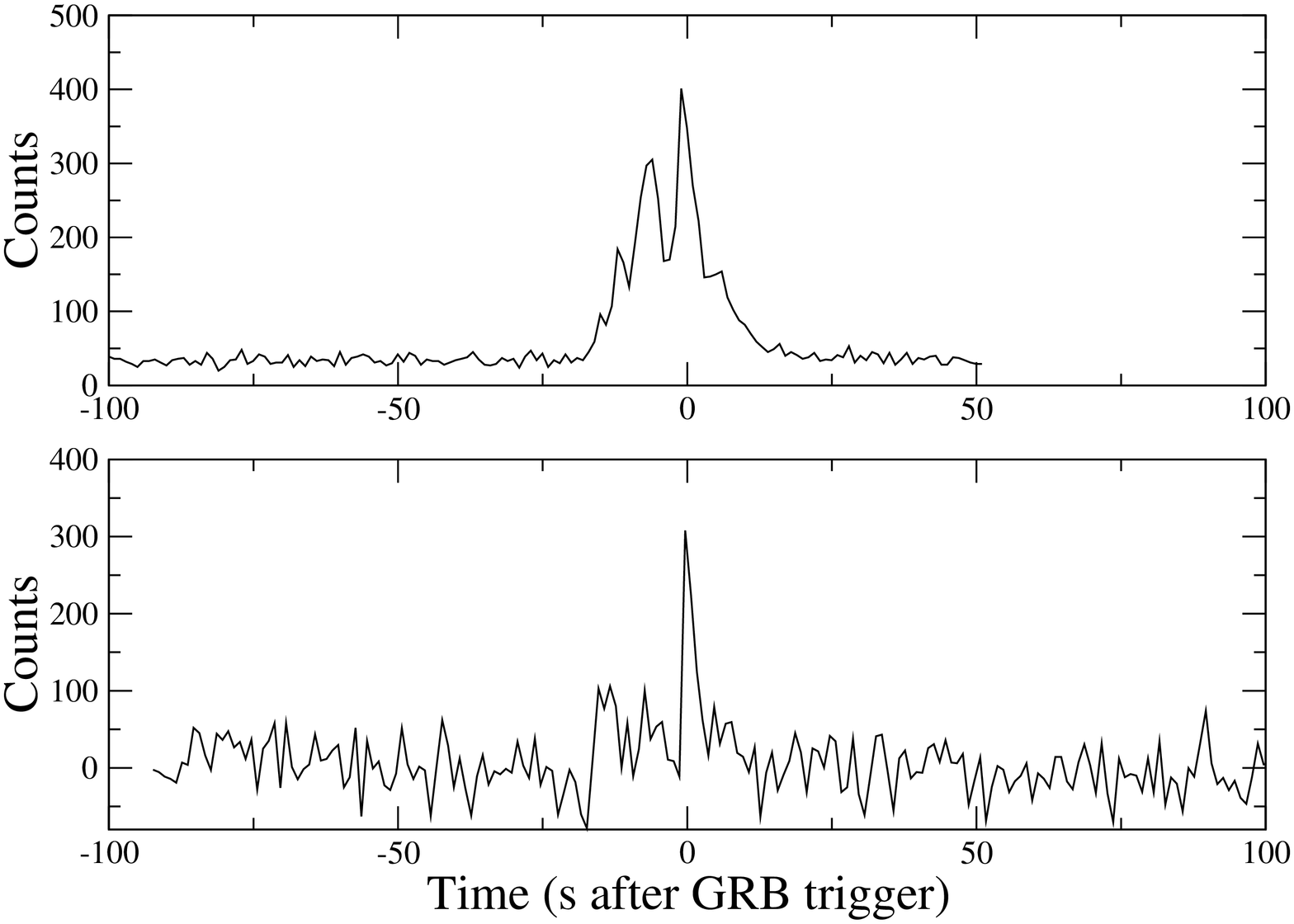}
                 \end{center}
         \end{minipage}
         &
         \begin{minipage}{0.48\textwidth}
                 \begin{center}
                 \includegraphics[height=6cm,angle=0]{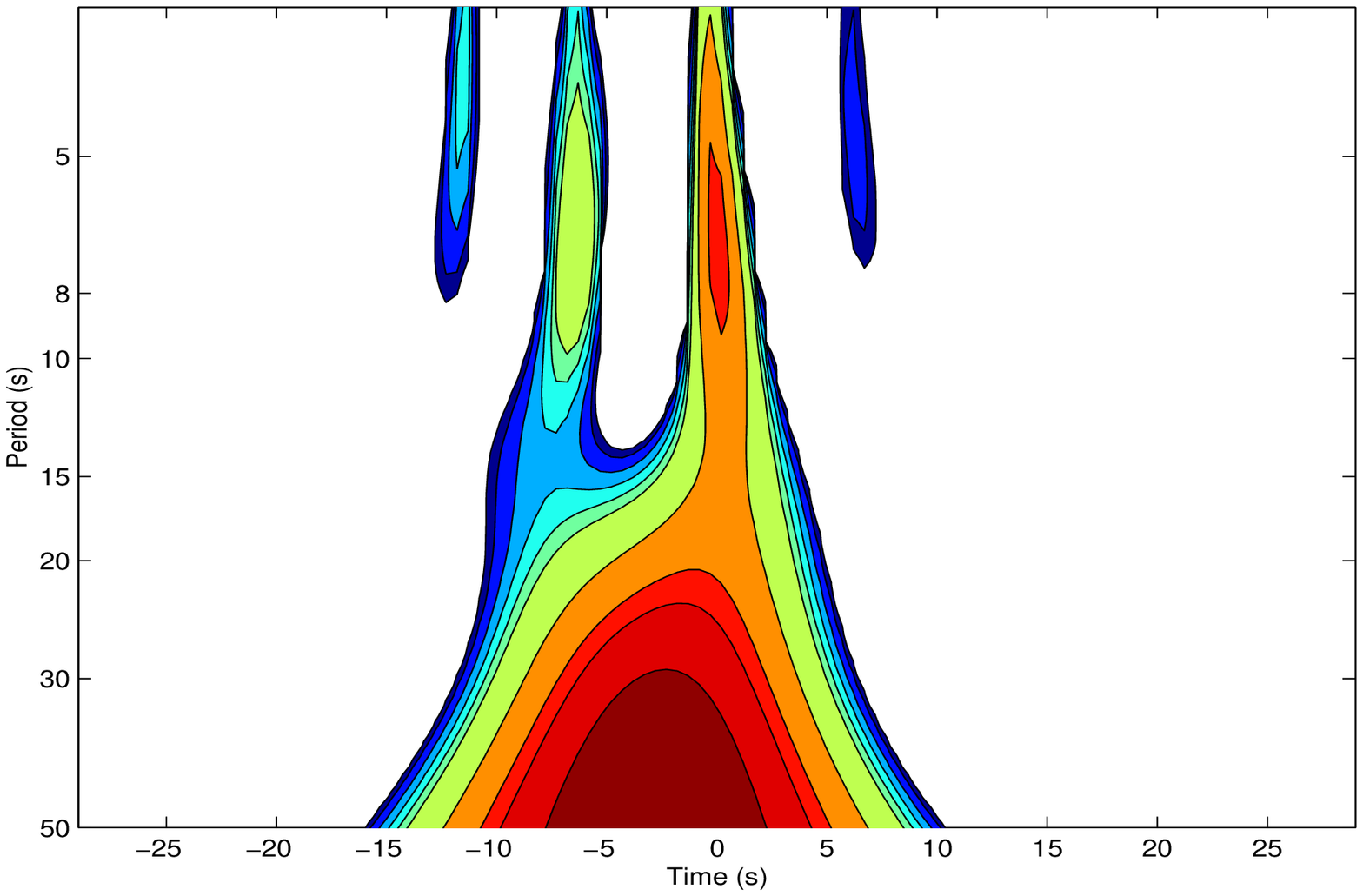}
                 \end{center}
         \end{minipage}
         \\
         \begin{minipage}{0.48\textwidth}
              \caption{\label{fig3}WFC (2--26 keV) and GRBM
          (40--700 keV) light curves of GRB~990704.}
        \end{minipage}
       &
         \begin{minipage}{0.48\textwidth}
          \caption{\label{fig4} WPAS of the original 10--26 keV light
             curve of GRB~990704. }
         \end{minipage}
 \end{tabular}
 \end{center}
 \end{figure*}

\begin{figure*}
 \begin{center}
\hspace{0.cm}
 \begin{tabular}{cc}
         \begin{minipage}{0.48\textwidth}
                 \begin{center}
                 \includegraphics[height=9cm,angle=0]{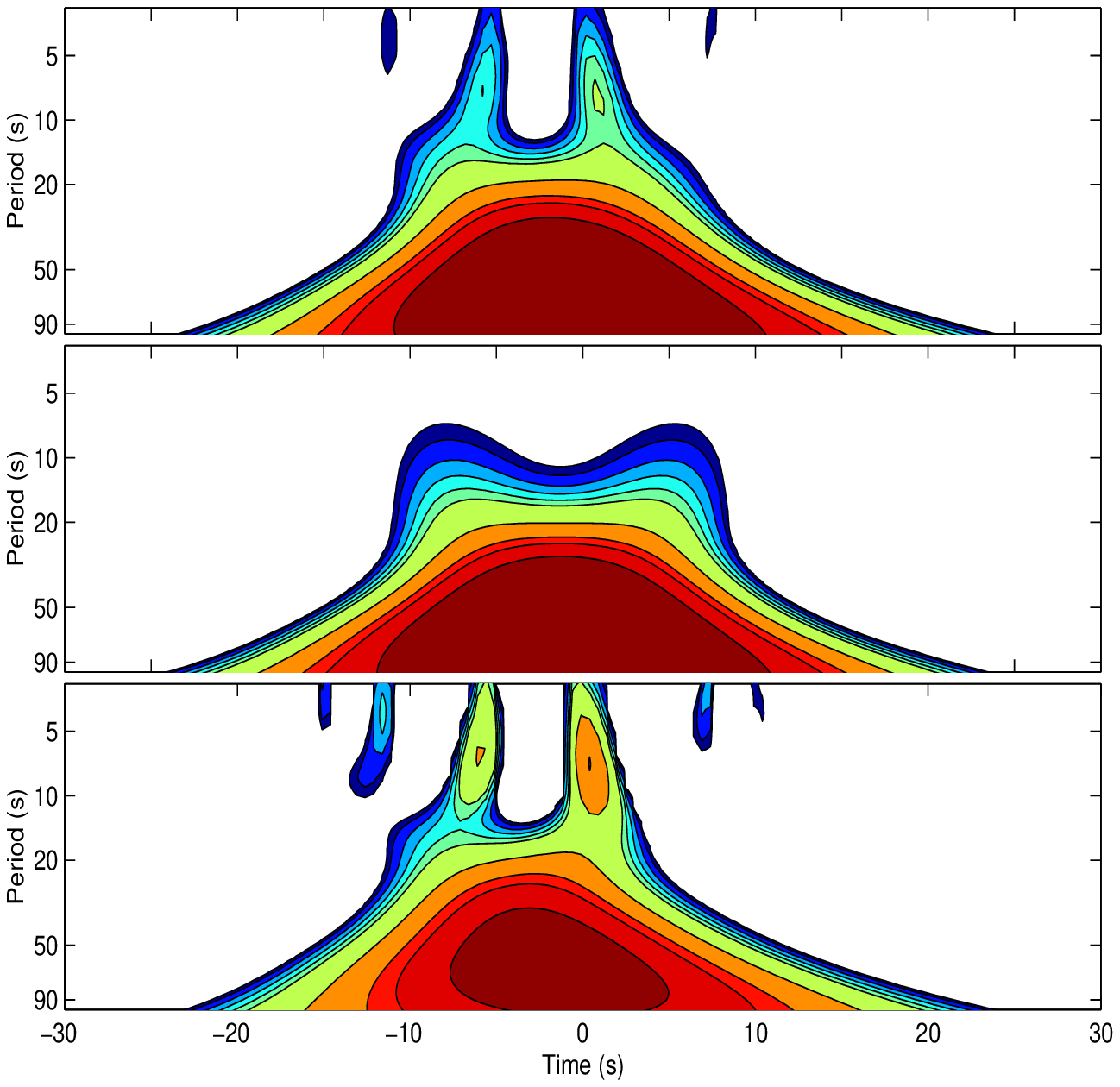}
                 \end{center}
         \end{minipage}
         &
         \begin{minipage}{0.48\textwidth}
                 \begin{center}
                 \includegraphics[height=9cm,angle=0]{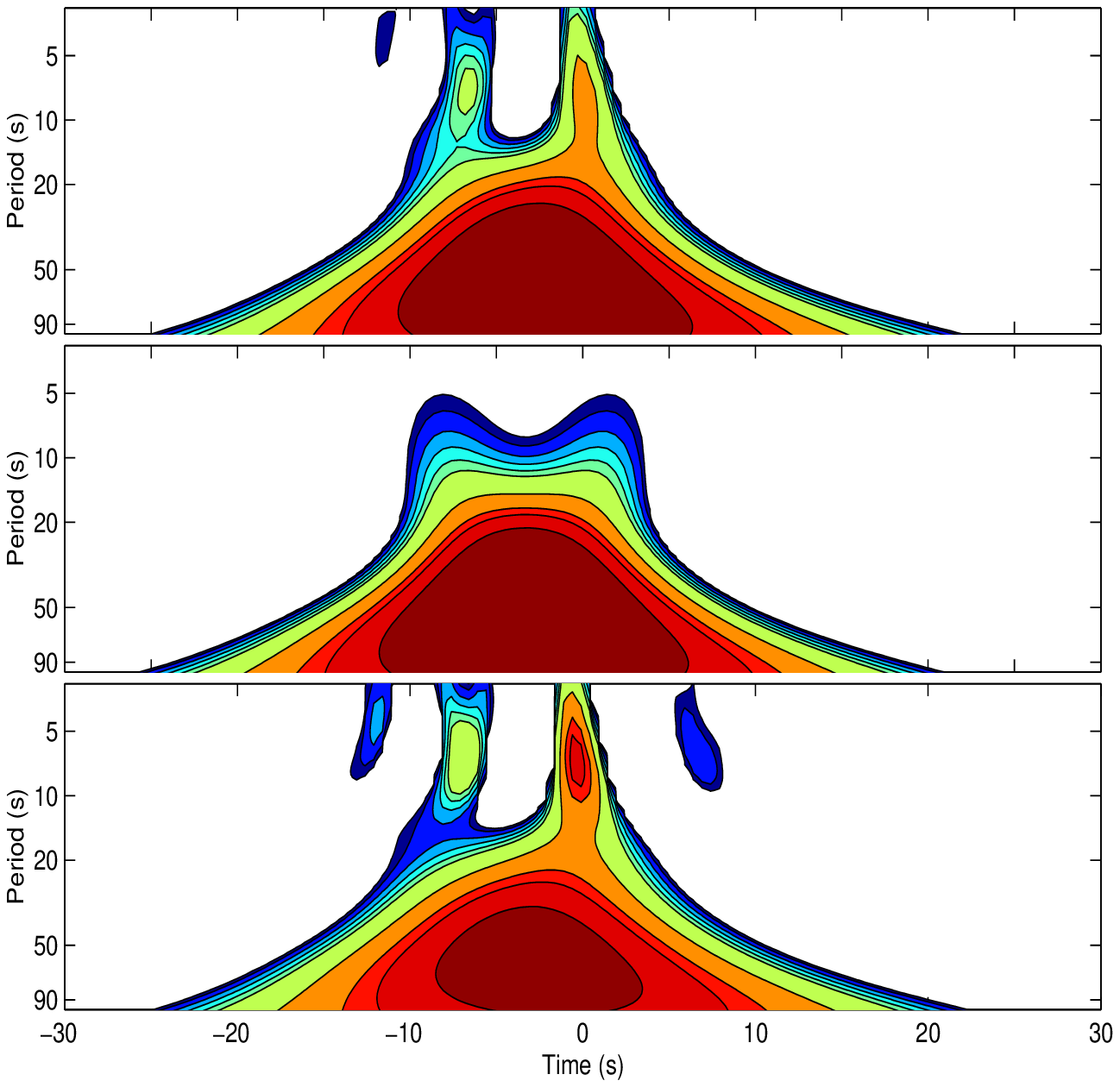}
                 \end{center}
         \end{minipage}
         \\
         \begin{minipage}{0.48\textwidth}
                \caption{\label{fig5} WPAS of the 2--5 keV light curve
                    of GRB~990704. Top panel: original light curve, middle
                    panel: SC, bottom panel: FC.}
        \end{minipage}
       &
         \begin{minipage}{0.48\textwidth}
          \caption{\label{fig6} WPAS of the 5--10 keV light curve of
                   GRB~990704. Top panel: original light curve, middle panel:
                   SC, bottom panel: FC.}
         \end{minipage}
 \end{tabular}
 \end{center}
 \end{figure*}

\subsubsection{\object{GRB~990704}}
On July 4, 1999 a GRB was detected by the \sax~ GRBM and WFC and
its position was given with a 7$'$ error radius (Heise et al.
1999a). A NFI observation, 8 hours after the burst, revealed an
X-ray afterglow (Feroci et al. 2001).
The X-ray emission shows a well apparent SC with some peaks
superposed, (Fig.~\ref{fig2}). This is a very interesting event
being the most X-ray rich \sax~ GRB: it is well apparent at 2--26
keV but only the highest peak is significantly detectable at
40--700 keV (Fig.~\ref{fig3}). The ratio of fluences in the WFC
and GRBM instrumental ranges is $F_{X}/F_{\gamma}=2.84\pm0.27$
(Feroci et al. 2001): the largest value found in the \sax~ sample
if we consider X-Ray Flashes as a separate class. The duration of
the event is strongly energy dependent, ranging from $\geq$ 30 s
in 2-5 keV to less than 20 s in the 40--700 keV energy range.

SC parameters were estimated from the 2--5 keV LC but,
after scaling the amplitude, we could not match the signal at 5--10
keV, because of the shorter duration. To find a curve able to
model the SC we considered first Eq.~(\ref{model}) and found a
satisfactory result only considering a symmetric curve (i.e. $b =
0$):
\begin{equation}
F_S(t,E_n)=A(E_n)~ exp[-C(E_n)~\left|(t-t_{0})\right|^s].
\end{equation}
where $C$ and $t_0$ were energy dependent to take into account the
change of SC profile. The comparison of the FCs with the light
curve at 10--26 keV shows a good accordance (Fig.~\ref{fig2}).
\begin{figure*}
 \begin{center}
\hspace{0.cm}
 \begin{tabular}{cc}
         \begin{minipage}{0.48\textwidth}
                 \begin{center}
                 \includegraphics[height=8cm,angle=-90]{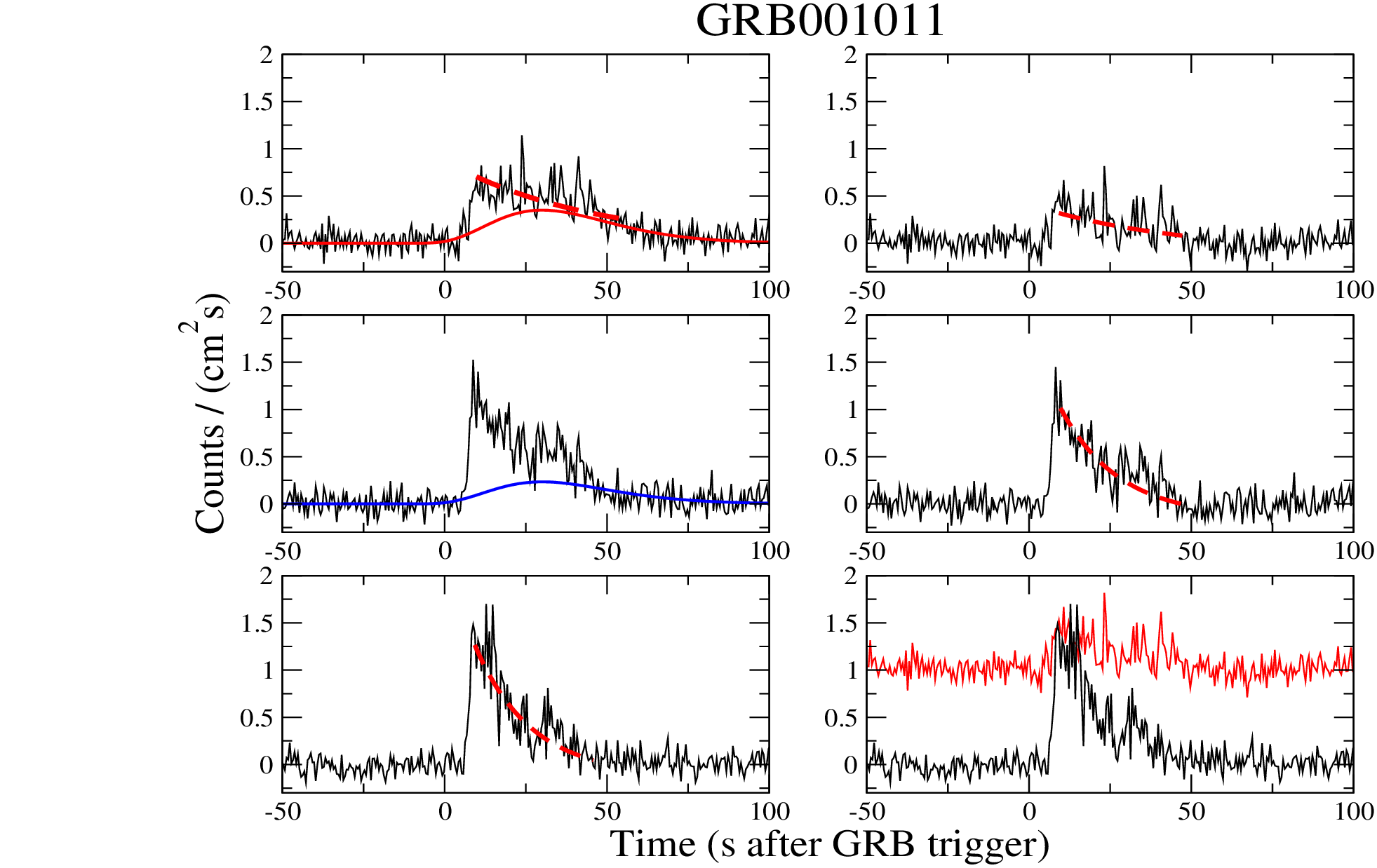}
                 \end{center}
         \end{minipage}
         &
         \begin{minipage}{0.48\textwidth}
                 \begin{center}
                 \includegraphics[height=8cm,angle=-90]{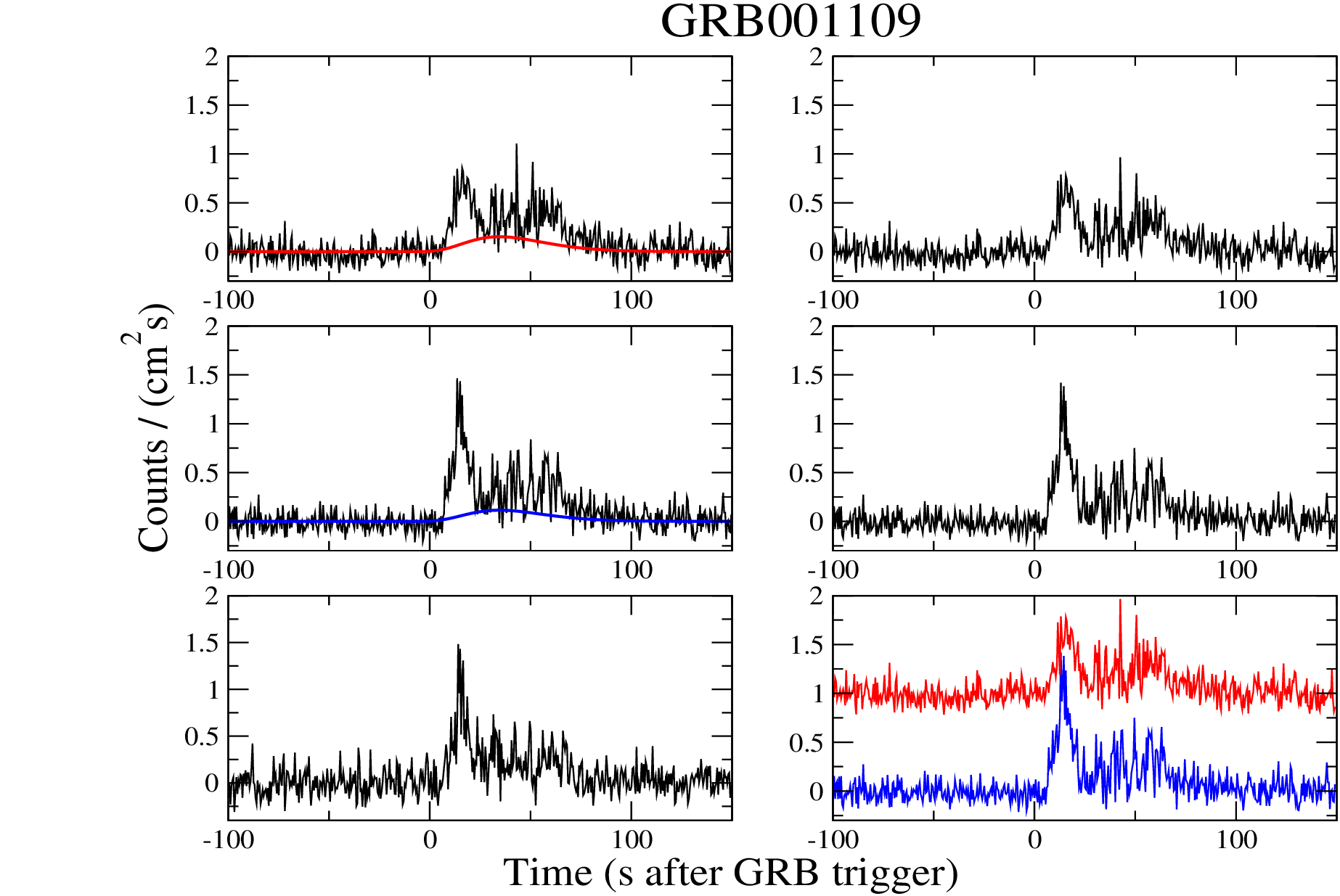}
                 \end{center}
         \end{minipage}
         \\
         \begin{minipage}{0.48\textwidth}
                \caption{\label{fig7}X-ray light curves of GRB 001011.
          Panel content is the same as Fig. 1. Dashed (coloured) lines
	  are exponential best fits of the decaying portion of LCs. }
        \end{minipage}
       &
         \begin{minipage}{0.48\textwidth}
          \caption{\label{fig8}X-ray light curves of GRB 001109.
        Panel content is the same as Fig. 1. }
         \end{minipage}
 \end{tabular}
 \end{center}
 \end{figure*}

A useful description of the power distribution over the various
time scales can be obtained from wavelet analysis.
We computed WPAS for all the LCs and their components. These are
shown in Fig.~\ref{fig5} (2--5 keV), Fig.~\ref{fig6} (5--10
keV) and Fig.~\ref{fig4} (10--26 keV): note that while in the
WPAS of the original data set (top panels) the power contribution
from narrow peaks (on time scales of a few seconds) is much less
apparent than that on longer times, being the signal dominated by SC,
the WPAS of FC (bottom panels) show clearly an enhanced relative power
from these features and compare well with the WPAS of the high energy
curve.
In particular, the WPAS in Fig. 4 shows how the power on time scales
shorter than $\sim$5 s is concentrated only in four features.
The two strongest central peaks are also present in the WPAS of
the original LC at lower energies (Figs. 5 and 6), while the others
become evident only in the WPAS of FCs. Note also that there is no
significant feature with a time scale between 10 and 30 seconds.

The complex time profile of this GRB is worthy of a more detailed
analysis.
Note first that the FC duration in the lowest energy range (Fig.~\ref{fig2},
upper right panel) is comparable to those in other ranges (see also
the WPAS in Figs. 5, 6). This implies that the longer duration
measured at 2--5 keV is essentially due to the SC.
Furthermore, it is clearly visible that the width of single peaks
decreases at higher energies and this effect continues to be apparent
even after the subtraction of the SC.
We verified it for the main peak of the FC (the highest at 5--10 keV)
and found that it agrees with Eq.~(\ref{Fenimore}).
We estimated for the two more prominent peaks the FWHM and found
$\Delta \tau(E_{1}) \simeq$ 3.5 s,
$\Delta \tau(E_{2}) \simeq$ 2.7 s and
$\Delta \tau(E_{3}) \simeq$ 1.9 s and their ratios,
when considering the central energies of our bands 
(i.e. 3.5, 7.5, 18 keV)
\begin{equation}\label{Dtau}
\frac{\Delta \tau(E_{i})}{\Delta \tau(E_{j})} \simeq
\Big( \frac{E_{j}}{E_{i}}\Big)^{0.45} 
\end{equation}
as expected from Eq.~(\ref{Fenimore}).

 In summary, when comparing the OR and the FC we find a similar energy 
behaviour of the main peaks, while the total prompt event duration  
is totally different. This finding implies that the SC subtraction  
does not modify the energy behaviour of the main features of the  
prompt event because the SC seems to be responsible only of the  
longer duration of the entire OR. This is clearly visible in  
Fig.~\ref{fig2} where the durations of the FC in all the three energy bands 
are just like that of the OR at 10-26 keV. 

As noticed above, the SC duration is also dependent on energy and in the
5-10 keV band it is significantly shorter than at 2-5 keV. Considering
the FWHM of their time profiles (20.7 s and 15.7 s at 2-5 keV and 5-10 keV,
respectively) as indicative of this difference we verified whether their 
ratio agrees with Eq.~(\ref{Dtau}). The resulting value of 1.41 
is indeed very close to that derived from the corresponding energies 
as in the case of the peaks of the FC.

 Though on different timescales, there is a common energy dependence 
between the SC and the features of the OR.  However it is still 
unclear what physics causes Fenimore's relationship, so it is unclear how
we can explain this partial similarity. 
Anyway, it is important to underline that this is the only 
burst to show this behaviour and it is an outstanding 
burst also for other properties. 

\subsubsection{\object{GRB~001011}}
This GRB was detected on 2000 October 11 at 15:54:50 UT
by the GRBM and WFC2 and localized with an accuracy of 5$'$
(Gandolfi et al. 2000a).
About 8 hours after the trigger, optical ($R$ band) and near
infrared ($J$ and $K_{s}$ bands) observations were performed at
the 1.54m Danish Telescope and the 3.58m New Technology Telescope,
respectively, both at ESO, La Silla, and led to the discovery
of a counterpart 
(Gorosabel et al. 2000, 2002).

WFC LCs have changing profiles at different energies.
In the 10--26 keV the time evolution is a typical FRED (Fast Rise
Exponential Decay) with a decay time of $\tau _{d}=$15.0 s.
In the 2--5 keV range it shows instead a more stable level, having
an approximate e-folding time $\tau _{d}=$ 44.8 s, with some
narrow peaks superposed.

\begin{figure*}
 \begin{center}
\hspace{0.cm}
 \begin{tabular}{cc}
         \begin{minipage}{0.48\textwidth}
                 \begin{center}
                 \includegraphics[height=8cm,angle=-90]{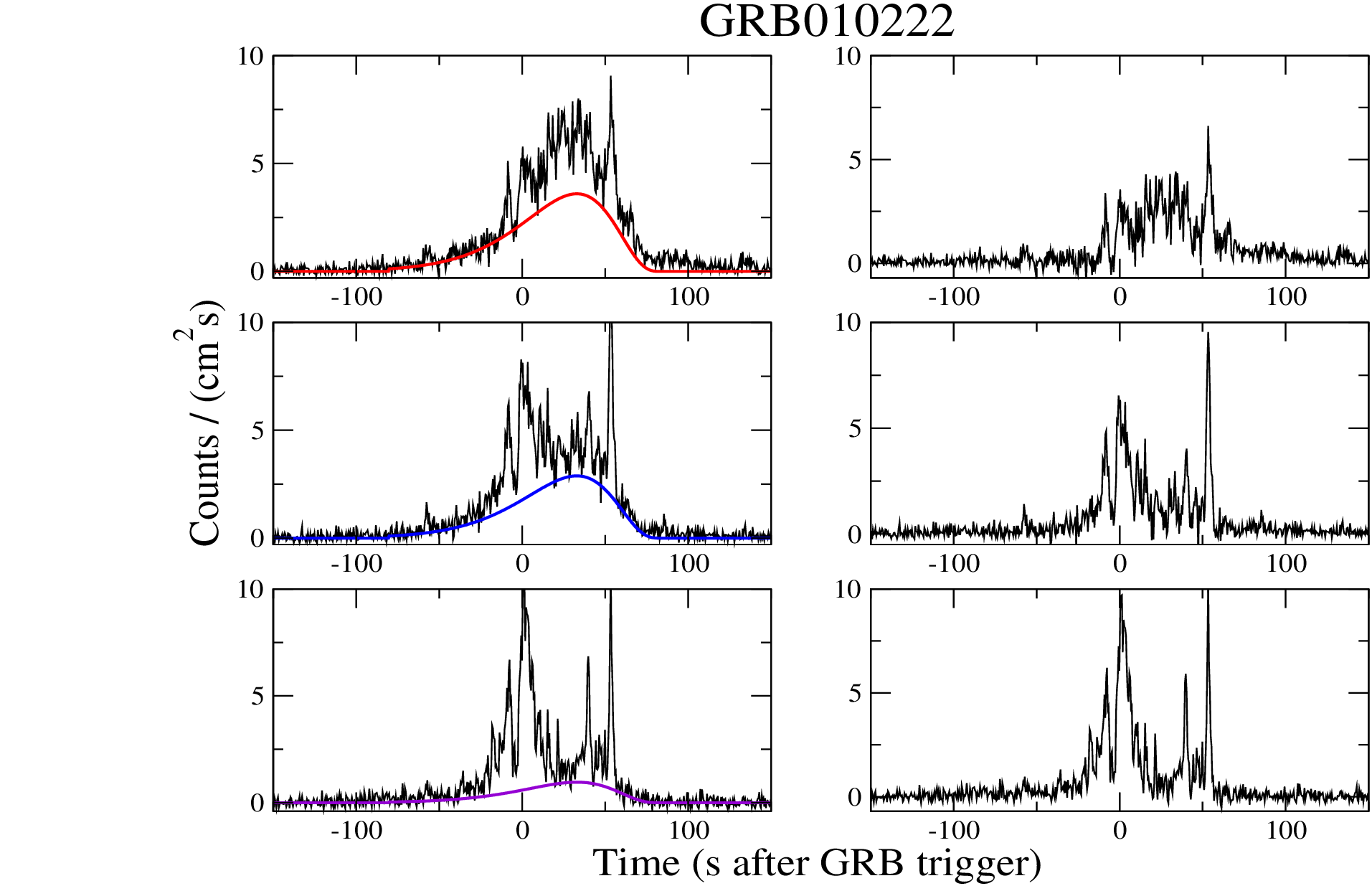}
                 \end{center}
         \end{minipage}
         &
         \begin{minipage}{0.48\textwidth}
                 \begin{center}
                 \includegraphics[height=9cm,angle=0]{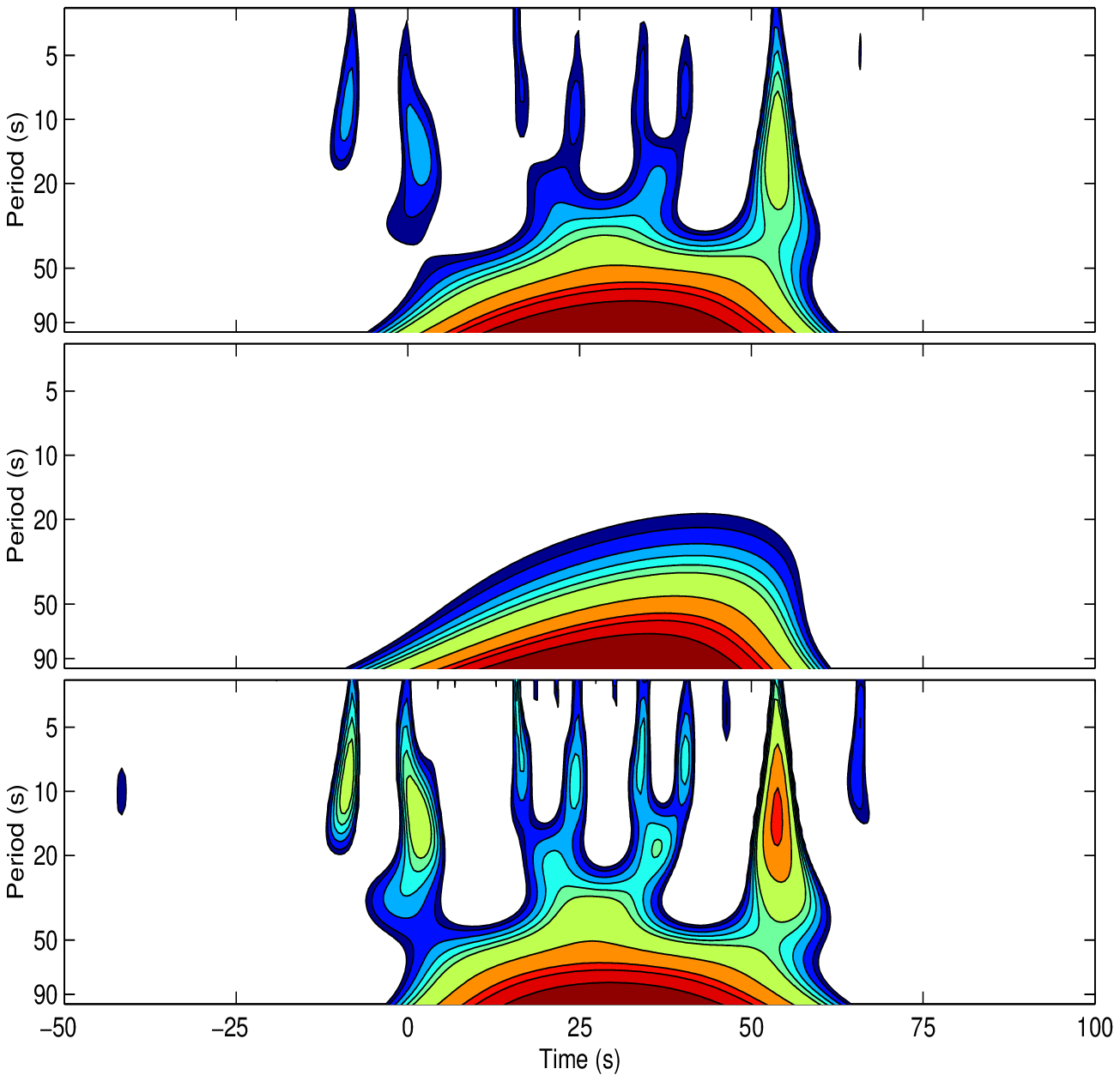}
                 \end{center}
         \end{minipage}
         \\
         \begin{minipage}{0.48\textwidth}
                \caption{\label{fig9}X-ray light curves of GRB~010222.
                  in the three considered energy ranges.
                  The panels on the left show the total counts
                  and the SC model; the panels on the right
                  show the corresponding FCs after the SC subtraction.}
        \end{minipage}
       &
         \begin{minipage}{0.48\textwidth}
          \caption{ \label{fig10} WPAS of the 2--5 keV light curve of
             GRB~010222. Top panel: original light curve, middle panel:
             SC, bottom panel: FC }
         \end{minipage}
 \end{tabular}
 \end{center}
\end{figure*}

\begin{figure*}
 \begin{center}
\hspace{0.cm}
 \begin{tabular}{cc}
         \begin{minipage}{0.48\textwidth}
                 \begin{center}
                 \includegraphics[height=9cm,angle=0]{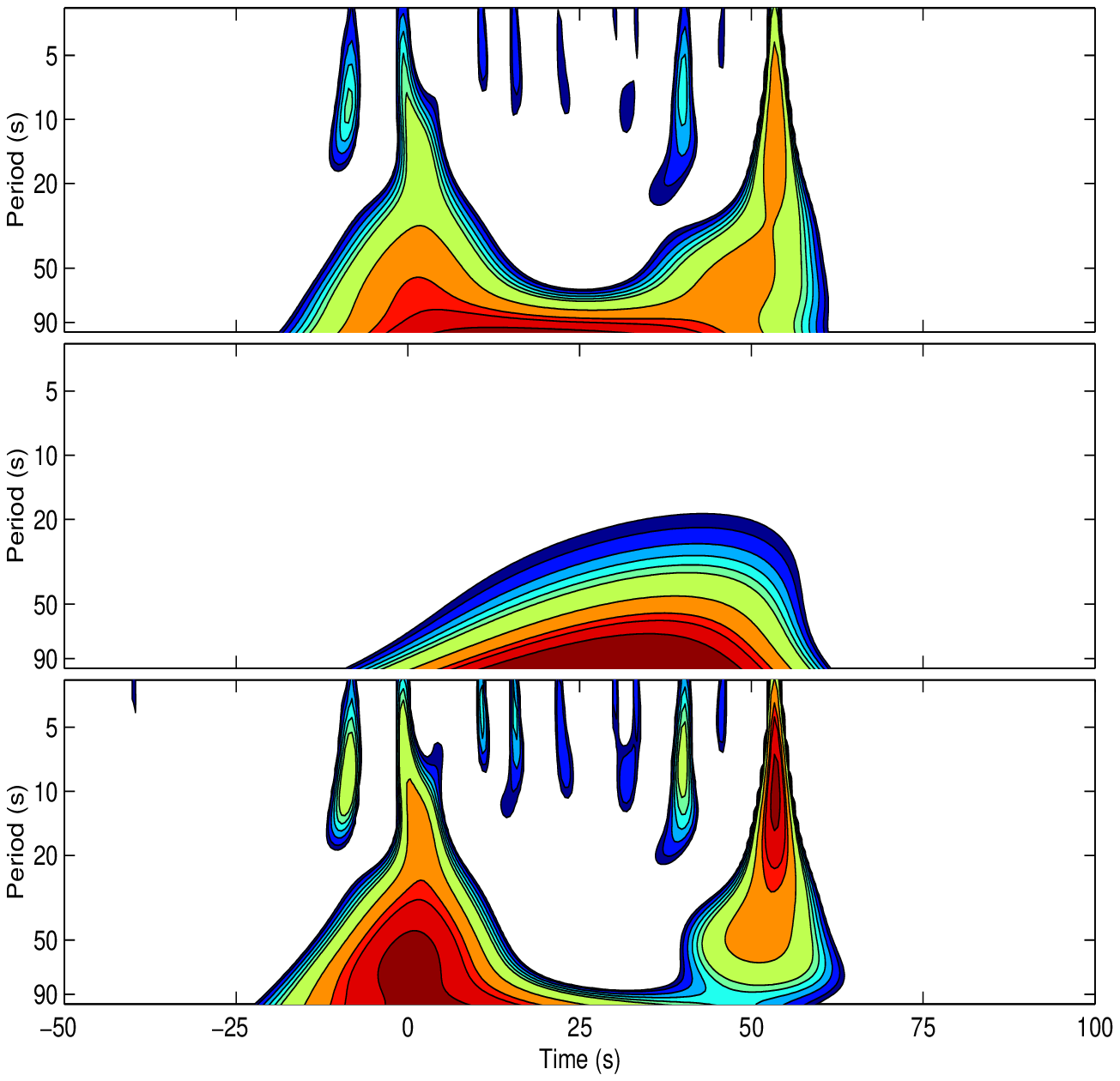}
                 \end{center}
         \end{minipage}
         &
         \begin{minipage}{0.48\textwidth}
                 \begin{center}
                 \includegraphics[height=9cm,angle=0]{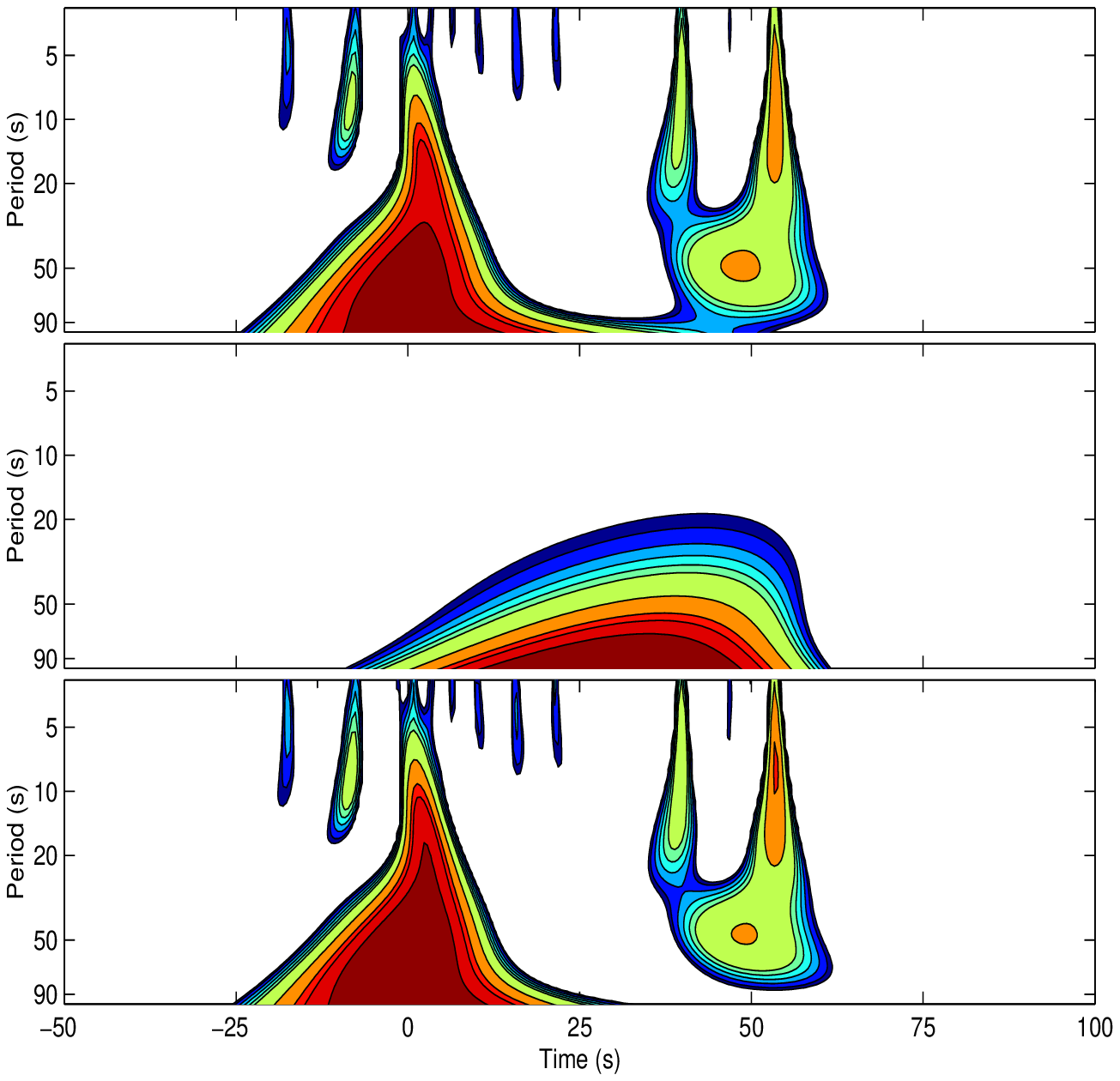}
                 \end{center}
         \end{minipage}
         \\
         \begin{minipage}{0.48\textwidth}
                \caption{\label{fig11} WPAS of the 5--10 keV light curve
                 of GRB~010222. Top panel: original light curve, middle
                 panel: SC, bottom panel: FC.}
        \end{minipage}
       &
         \begin{minipage}{0.48\textwidth}
          \caption{ \label{fig12}  WPAS of the 10--26 keV light curve
                 of GRB~010222. Top panel: original light curve, middle
                 panel: SC, bottom panel: FC.}
         \end{minipage}
 \end{tabular}
 \end{center}
\end{figure*}

In this case it was not easy to find the right curve to represent
the SC. A possible model (2--5 keV) is shown in the upper panel
(left side) of Fig.~\ref{fig7}: it is nearly symmetric with a
rather mild rising portion followed by a slower decay. As before,
the SC at 5-10 keV has a smaller amplitude, a decay time $\tau
_{d}=$23.5 s, while it disappears at 10-26 keV (Fig.~\ref{fig7},
right panels). We used the 2-5 keV LC to estimate the parameters
of the SC and the resulting FC is plotted in the same figure
(upper panel, right side), which has practically the same decay
time of the original LC.
We fitted the same SC function to the 5--10 keV LC, changing only
the normalization, and found a decay time of the FC equal to 18.1
s, well comparable with that derived for the 10-26 keV curve
(Fig.~\ref{fig7}, left panels). Taking into account these
differences in the decay times of the LCs above $\sim$5 keV with
respect to that at lower energies, we cannot exclude that the SC
has a very soft spectrum that makes it apparent only in the 2--5
keV LC.

\subsubsection{\object{GRB~001109}}
On November 09, 09:23:17 UT this GRB was detected by the GRBM and
WFC1 with a refined position uncertainty of 2.5$'$ (Gandolfi et
al. 2000b,c).
The NFI X-ray follow up showed a previously unknown source
(Amati et al. 2000a).
Other observations in the radio (Taylor et al. 2000, Berger \&
Frail 2001a) and in the optical-near infrared (Castro Cer\'on et
al. 2004) showed sources which could not be related with
\object{GRB~001109}.

The 2--5 keV LC (Fig.~\ref{fig8}, upper left panel) is
characterised by a prominent peak about at the beginning of the
prompt emission followed by others of smaller amplitudes. All these peaks
are superposed on a small amplitude SC, which is further reduced
at 5--10 keV and undetectable at 10--26 keV. As before we
estimated the SC parameters from the low energy LC and the
amplitude was scaled to derive the SC at 5--10 keV. As in other
cases we found a good accordance between the 10--26 keV LC and the
estimated FCs at 2--5 keV and 5--10 keV (Fig.~\ref{fig8}, right panels). 
Note,
however, that together with \object{GRB~980519} the SC of this burst has a
low fractional content. Furthermore, it shows a prominent
symmetric initial peak whose FWHM is clearly increasing with
energy.

\subsubsection{\object{GRB~010222}}
GRB~010222 was detected on 2001 February 22 at 07:22:34.9 UT
by GRBM and WFC1 and its position was determined with an accuracy
of 2$'$.5 (Piro 2001a).
Only  3.2 h after the burst an alert message was distributed so
follow-up observations were performed very quickly.
The discoveries of an optical and radio counterparts were announced
4.4 h after the burst (Henden 2001a, 2001b) and 7.7 h (Berger
\& Frail 2001b), respectively.
Other detections followed in the $R$ band (Stanek et al. 2001a),
near-infrared (Di Paola et al. 2001), and at sub-mm wavelengths
(Fich et al. 2001).

\object{GRB~010222} is a very strong burst and the count rates in all the
X-ray bands are a factor of about four higher than for other GRBs.
The isotropic energy output was estimated
$E_{iso}$=(154.2$\pm$17)$\times$10$^{52}$ erg (Amati et al. 2002)
at a redshift $z$ = 1.477 (Stanek et al 2001b).
The 2--5 keV LC shows several peaks superposed on an evolving SC
which at 5--10 keV has a comparable amplitude and it is strongly
reduced at 10-26 keV.
Likely,   an SC in this energy band would not be apparent
if the count rate had a S/N ratio like those of other GRBs.

The time profile of SC is anomalous with respect other GRBs because
it has a slow rise followed by a faster decay. We then preferred
to swap in time the model equation:
\begin{equation}
F_S(t,E_n)=[A(E_n)]~ (t_{0}-t)^b~ exp[-C~(t_{0}-t)^s]
\end{equation}
where $t_{0}=163.6$ s. SC parameters were estimated from the 2--5
keV LC and the same function was used to fit the SC at higher
energies leaving free only the normalization. The resulting FCs
are shown in the right panels of Fig.~\ref{fig9} and present
similar structures indicating that SC was well modeled.

Also for this GRB we studied the time structure of LCs by means
of wavelet analysis and the WPAS confirm the components' behaviour.
Fig.~\ref{fig10} shows that before the subtraction of the SC the
power on long time scales was not symmetrically distributed with
respect to the central time of the 2--5 keV LC.
This effect is due to the SC which has the maximum at about $t$ =
30 s, starting from the trigger instant. 
After the SC subtraction the power of the resulting FC
does not continue to show this asymmetry and the power content
on much shorter time scales is clearly apparent.
WPAS of the FCs at 5--10 and 10--26 keV (Fig.~\ref{fig11},~\ref{fig12})
looks nearly identical with a series of features proceeded by the narrow
peaks.
In particular, note also for this GRB, the lack of power on time
scales longer than 5 s in the central portion of the GRB.

Finally, it is important to note that for this burst the SC is
much longer than FC, particularly on the leading side and this is
clearly apparent in the LCs of Fig.~\ref{fig9}: the
series of narrow peaks of the FC starts about 20 s before the
trigger time, while the smooth increase of the SC begins about
90 s before.

\subsubsection{\object{GRB~010412}}
This GRB was detected on 2001, April 12 at 21:46 UT by the GRBM
and WFC1 (Piro 2001b). There was no follow-up observation with ~\sax 
NFI was scheduled due to an ongoing TOO.
No previous analysis and information on a possible afterglow
can be found in the literature.

\begin{figure*}
 \begin{center}
\hspace{0.cm}
 \begin{tabular}{cc}
         \begin{minipage}{0.48\textwidth}
                 \begin{center}
                 \includegraphics[height=8cm,angle=-90]{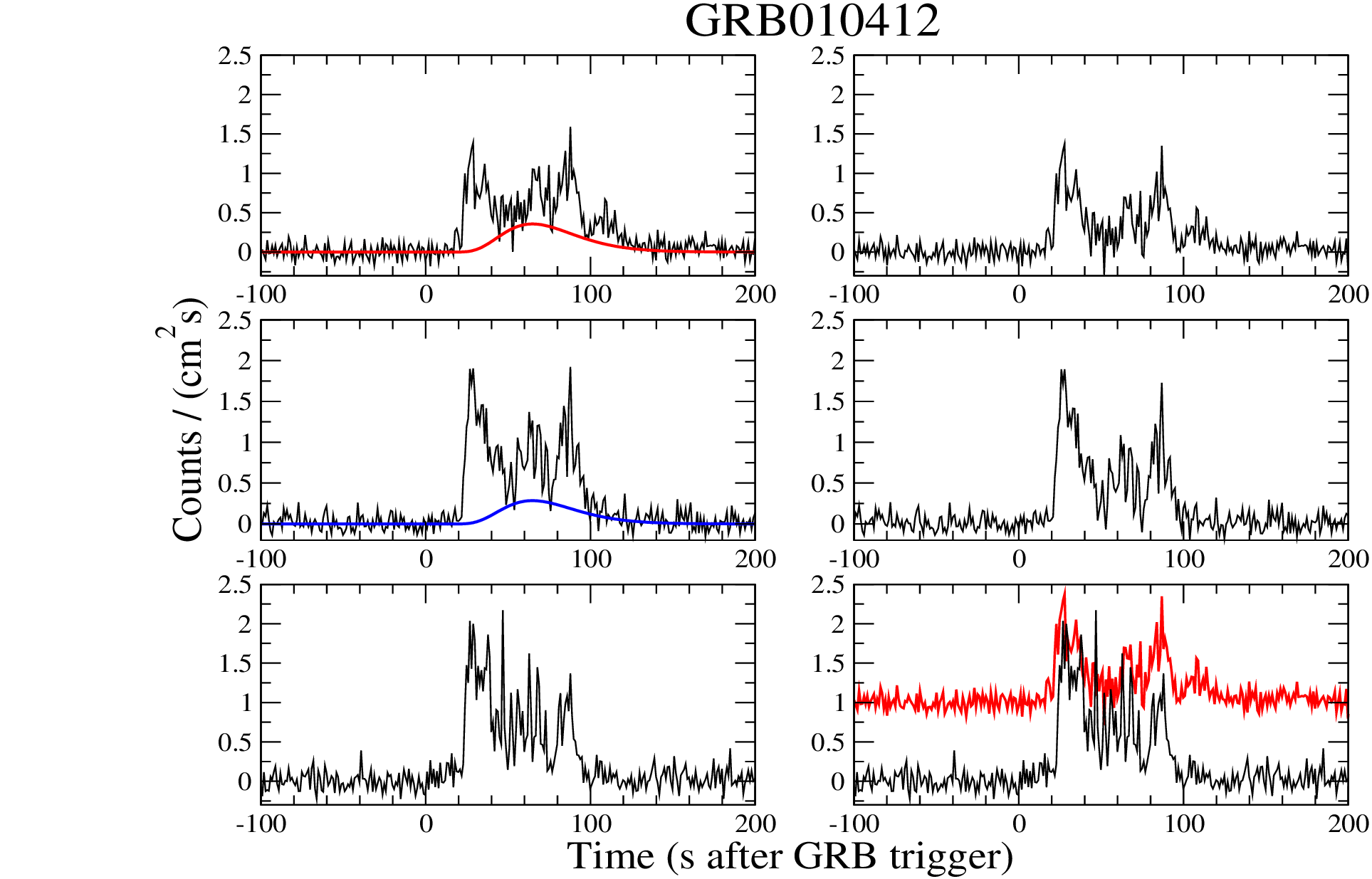}
                 \end{center}
         \end{minipage}
         &
         \begin{minipage}{0.48\textwidth}
                 \begin{center}
                 \includegraphics[height=8cm,angle=-90]{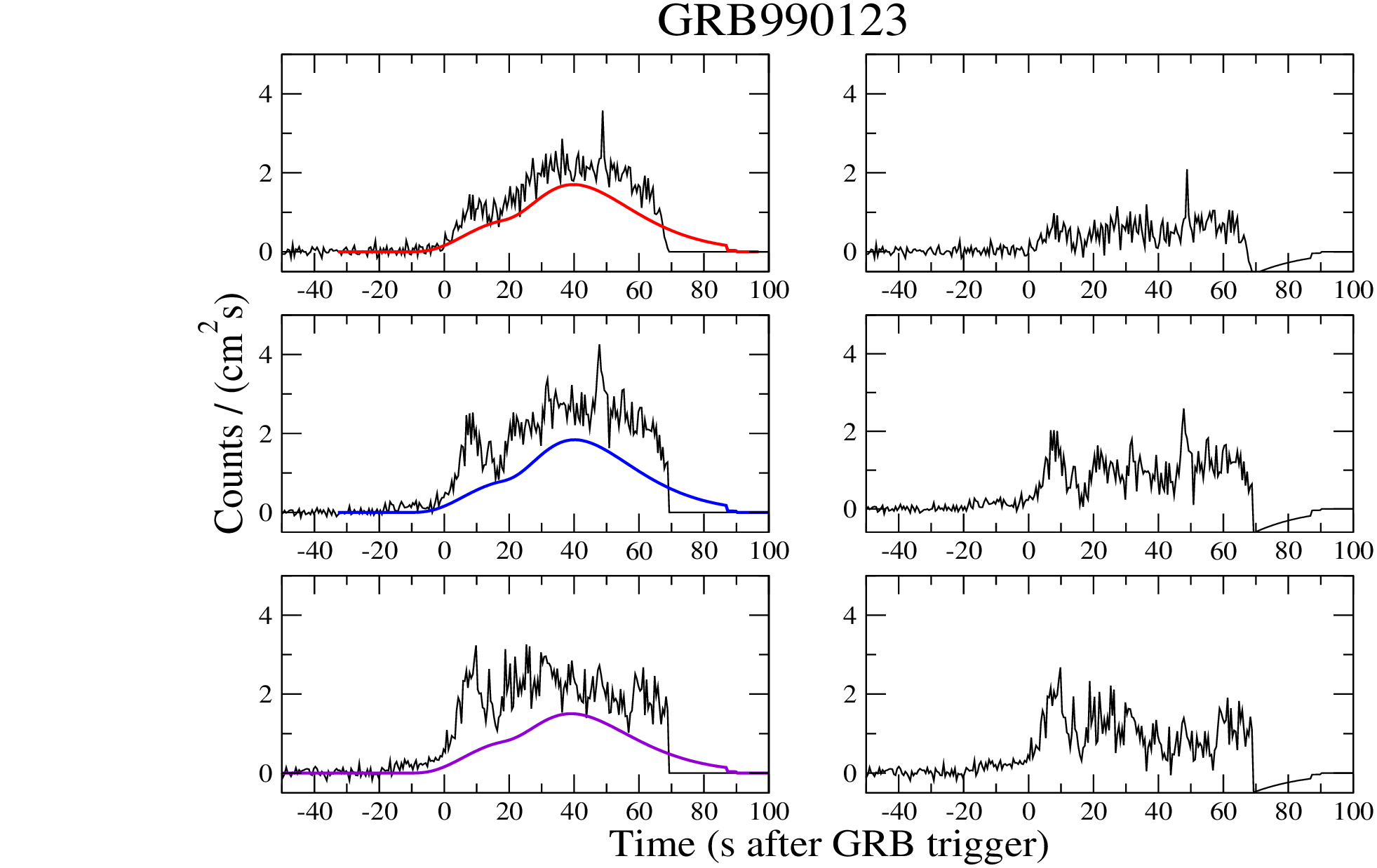}
                 \end{center}
         \end{minipage}
         \\
         \begin{minipage}{0.48\textwidth}
                \caption{\label{fig13}X-ray light curves of GRB 010412.
          Panel content is the same as Fig. 1.}
        \end{minipage}
       &
         \begin{minipage}{0.48\textwidth}
          \caption{\label{fig14}X-ray light curves of GRB 990123.
          Panel content is the same as Fig. 1. }
         \end{minipage}
 \end{tabular}
 \end{center}
 \end{figure*}

The LCs of this event are characterised by two main peaks at the
beginning and the end of the prompt emission with several smaller and
narrower peaks between them. A SC is present in the 2--5 and 5--10
keV curves but it not apparent at higher energies. Also in this
case we estimated the SC from the lowest energy data and found a
satisfactory agreement between the resulting FCs with the LC in
the highest energy band (Fig.~\ref{fig13}).
A residual SC with a very small amplitude can be seen at 10--26 keV,
but it could be an artifact due to peak blending.

\subsection{GRBs with hard Slow Components}

\subsubsection{\object{GRB~990123}}
\object{GRB~990123} was detected by WFC1 on January 23 at 09:47:13.9 UT.
It is the only burst whose prompt emission was detected in X- and 
$\gamma$-ray bands and also in the optical (Akerlof et al. 1999a, 1999b, 
Feroci et al. 1999).
Its afterglow was followed in the X rays (Heise et al. 1999b), optical
(Odewahn et al. 1999), and radio (Frail et al. 1999) after the rapid
localization (Piro et al. 1999a, 1999b, 1999c) so that its spectrum
and evolution were extensively studied (Galama et al. 1999a, Fruchter 
et al. 1999, Kulkarni et al. 1999a, Sari \& Piran 1999, Corsi et al.
2005).

Optical observations began with the ROTSE-I telephoto array
while the $\gamma$-ray event was still in progress: the
optical transient (OT) reached a peak about 32 s after the
start of the burst with 8.9 mag (Akerlof \& McKay 1999).
Within hours, spectroscopy revealed metal absorption lines in
the spectrum of the OT at $z$ = 1.60 (Kelson et al. 1999; Hjorth
et al. 1999a).
This was followed by a detection of a strong unusual radio
flare from the same source (Kulkarni et al. 1999; Galama et al.
1999).

\begin{figure*}
 \begin{center}
\hspace{0.cm}
 \begin{tabular}{cc}
         \begin{minipage}{0.48\textwidth}
                 \begin{center}
                 \includegraphics[height=8cm,angle=-90]{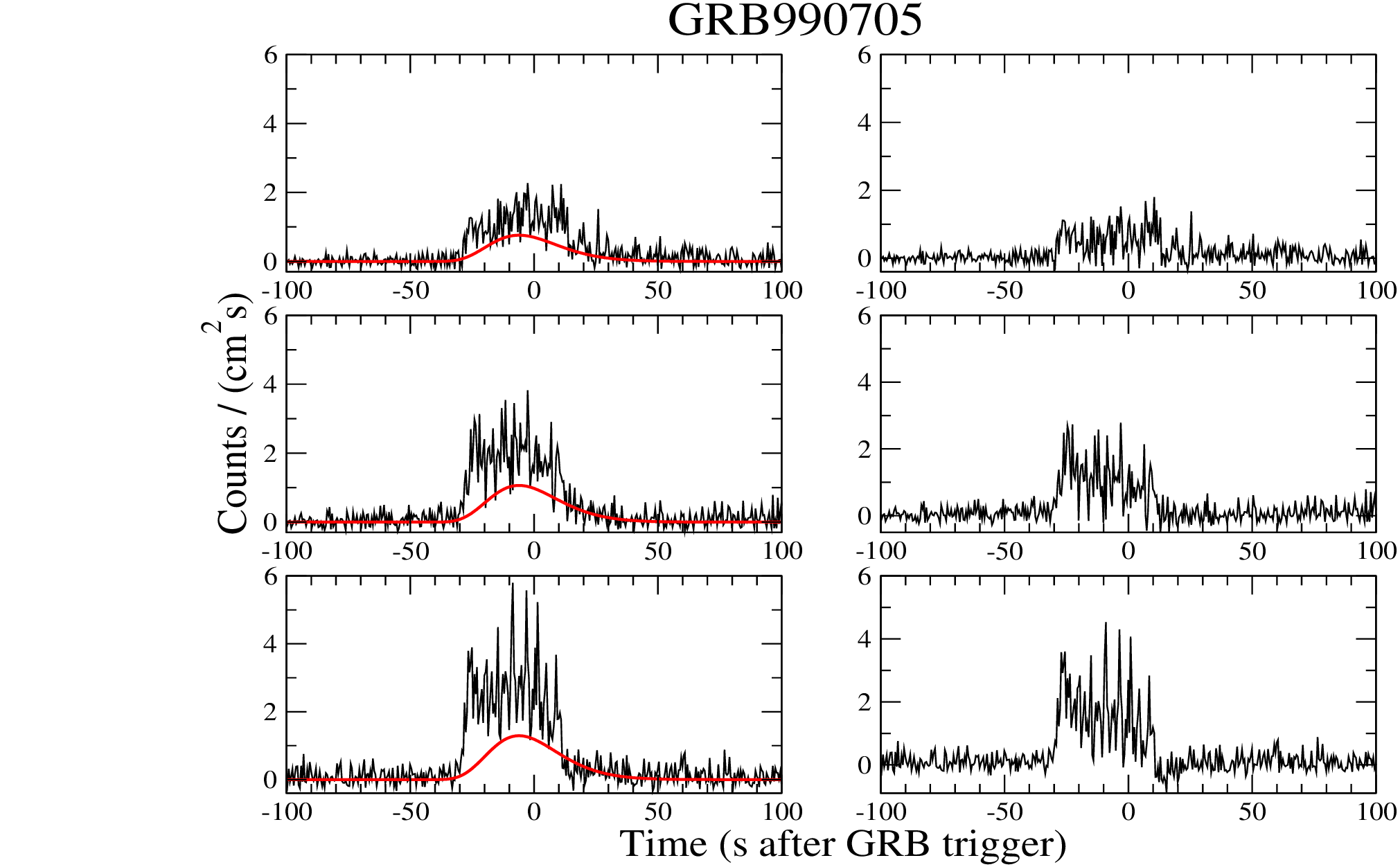}
                 \end{center}
         \end{minipage}
         &
         \begin{minipage}{0.48\textwidth}
                 \begin{center}
                 \includegraphics[height=8cm,angle=-90]{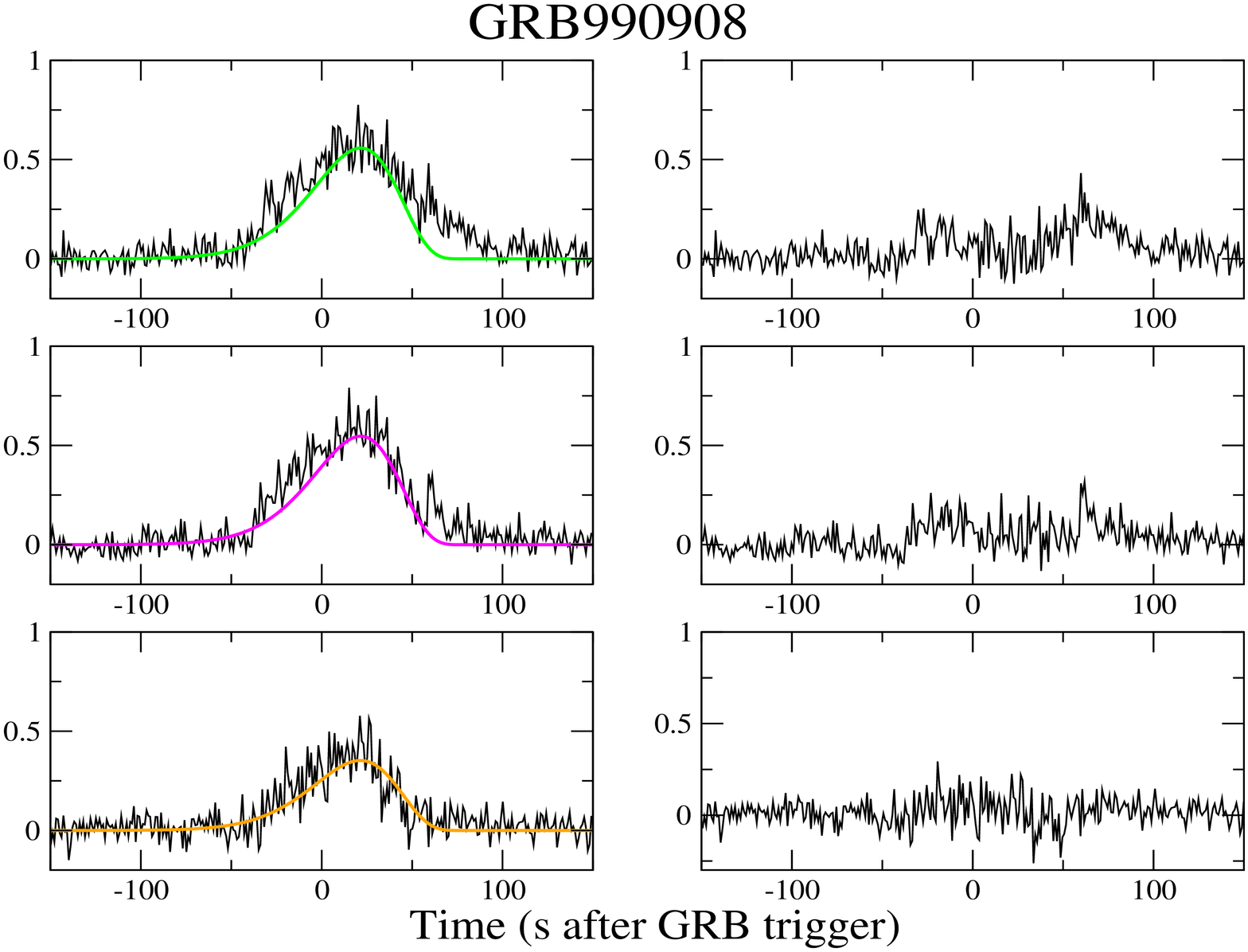}
                 \end{center}
         \end{minipage}
         \\
         \begin{minipage}{0.48\textwidth}
                \caption{\label{fig15}X-ray light curves of GRB 990705.
        Panel content is the same as Fig. 9.}
        \end{minipage}
       &
         \begin{minipage}{0.48\textwidth}
          \caption{\label{fig16}X-ray light curves of GRB 990908.
        Panel content is the same as Fig. 9. }
         \end{minipage}
 \end{tabular}
 \end{center}
 \end{figure*}

\object{GRB~990123} is the strongest burst detected by the WFCs with the
isotropic energy output $E_{iso}$ = (278$\pm$32)
$\times$10$^{52}$ erg (Amati et al. 2002).
The 2--5 keV LC is characterized by the presence of a relevant SC
with {\bf a} growing amplitude at 5-10 keV and at 10-26 keV (Fig.~\ref{fig14}).
We were not able to model an acceptable SC pattern using only
Eq.~(\ref{FC}) and reached a satisfactory result considering the sum
of two similar functions. The parameters' values for the two model
functions are given separately in the two lines in Table 2.
Note that the WFC light curves are not complete because the burst
direction was obscured by Earth atmosphere and, consequently, the
evolution of the X-ray SC in the last part of the event is unknown.

This complex SC modelling required different amplitude ratios for the
two functions. We preferred to change only the amplitude of the second
function, which varied of about 15\%, and to keep that of the first
function constant, although it could be increasing at higher energies.
Another problem in the SC modelling arises from the different spectra
of short peaks. For instance, there is a strong peak at the beginning
of the  prompt emission which becomes more apparent at higher 
energies, whereas
the spike at $t=32$ s, well visible at 2--5 keV is smaller and broader
at 5--10 keV and practically disappears above 10 keV. Note that this
behaviour is opposite to that of the typical peak described by the
Eq.~(\ref{Fenimore}).

Time resolved X-ray spectroscopy of this GRB was recently performed
by Corsi et al. (2005) without a distinction between possible slow
and fast components . These authors find that spectra cannot be
well represented by a single simple law from the optical to X and
$\gamma$ rays and suggest that emission mechanisms working at low and
high frequencies may be different (see also Vestrand et al. 2005). 
We will discuss better this result  in Sect. 5.

\subsubsection{\object{GRB~990705}}

It was detected by GRBM and WFC on July 5, 1999  at 16:01:25 
(Celidonio et al. 1999).
It is the second brightest in $\gamma$-rays (40--700 keV) after \object{GRB
990123} and ranks in the top 15\% in X-rays (2--26 keV).
Optical and
near-infrared observations of the GRB~990705 location led to the
discovery of a reddened fading counterpart and a possible host
galaxy (Masetti et al. 2000). Holland et al. (2000) have imaged
the \object{GRB~990705} field with the Hubble Space Telescope, detecting a
spiral galaxy at the GRB position: although the distance of this
galaxy is not known, its size and brightness were found compatible
with a redshift $z \leq$ 1. A redshift estimate at $z=0.86$ has
been derived by Amati et al. (2000b) from an absorption edge in
the X-ray spectrum. \object{GRB~990705} is characterised by an isotropic
energy output $E_{iso}$ = (3.5$\pm$0.15)$\times$10$^{53}$ erg
(Amati et al. 2002).

The X-ray LCs of \object{GRB~990705} suggest the presence of a SC whose
amplitude is increasing with energy. We applied to them our
standard analysis using Eq.~(\ref{model}) and and obtained a nearly
symmetric SC whose amplitude ratios $1 < k_{21} < k_{31}$ confirm
its hard spectrum (Table 2). 
Fig.~\ref{fig15} shows the X-ray curves of SCs and of the resulting
FCs. The latters show only one large feature at the beginning of 
the burst followed by a series of much narrower peaks.

\subsubsection{\object{GRB~990908}}

This GRB was detected on September 8, 1999 at 00:18 UT
by GRBM and WFC instruments with a position accuracy of 8$'$ 
(Piro et al. 1999d).
No subsequent \sax~ observation was
performed, because the follow-up of \object{GRB~990907} was still ongoing
and, therefore no information on a possible afterglow is
available. There are no published data on this GRB. The average
count rate is lower than other bursts and LCs appear noisy:
their similar shapes in the three bands indicate that SC is dominating
over the entire duration of about 150 s (Fig.~\ref{fig16}).
Only a few narrow peaks are barely detectable at the beginning and
at the end of LCs indicating that FCs are much weaker than other GRBs,
if present at all.
The 2--5 keV SC has quite similar rise and decay times, whereas
at higher energies the latter seems shorter. To reproduce this
behaviour we inverted the time direction in the standard
SC modelling. Parameters were evaluated from the 5--10 keV LC
and then the same function was adapted to the other bands.
The result is acceptable although it is not very precise as
apparent from the residual FCs shown in Fig.~\ref{fig16}
(left panels).

\begin{figure}
      \hspace{0cm}
\includegraphics[height=8.cm,angle=-90]{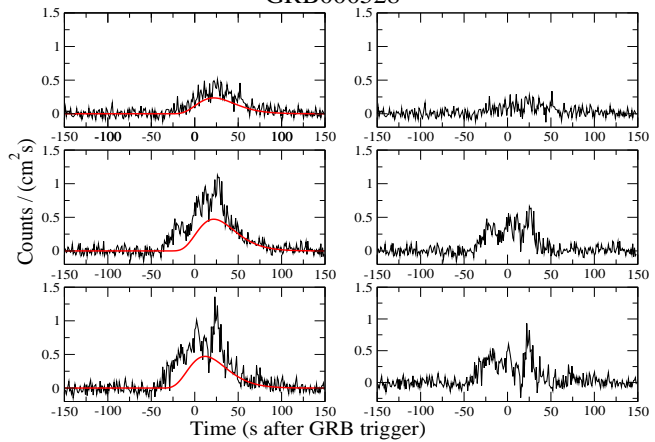}
\caption{X-ray light curves of GRB 000528.
        Panel content is the same as Fig. 9.}
\label{fig17}
\end{figure}

\begin{table*}
\centering
\caption{Integrated counts (N$_i$) and SC fractional content (f$_i$)
 in the three WFC energy bands and in the total 2--26 keV range 
(N$_{tot}$, f$_{tot}$).  Numbers in brackets are obtained using 
the SC values only in the first two ranges.} \label{tab3}
\begin{tabular}{ccccccccc}
\hline GRB & $N_1$ & $f_1$ & $N_2$ & $f_2$ & $N_3$ & $f_3$ &
$N_{tot}$ & $f_{tot}$ \\
    & ct/cm$^2$ &  & ct/cm$^2$ &  & ct/cm$^2$ &  & ct/cm$^2$ &  \\
\hline \object{GRB 980519} & 87.2 & 0.27 & 81.7 & 0.17 & 51.9 &  & 216.3 &
(0.17) \\ \object{GRB 990704} & 66.7 & 0.66 & 53.2 & 0.45 & 26.6 &  & 146.5 &
(0.46) \\ \object{GRB 001011} & 24.9 & 0.62 & 27.6 & 0.37 & 21.0 &  & 73.5 &
(0.35) \\ \object{GRB 001109} & 25.9 & 0.28 & 26.6 & 0.20 & 20.4 &  & 72.9 &
(0.17) \\ \object{GRB 010222} & 441.6& 0.57 & 392.3& 0.52 & 277.9& 0.25 &
1111.8 & 0.47\\ \object{GRB 010412} & 61.9 & 0.31 & 71.7 & 0.21 & 64.1 &  &
197.7 & (0.17) \\ \hline
\object{GRB 990123} & 104.3& 0.67 &140.3 & 0.53 &132.8 & 0.48 & 377.4 & 0.56 \\
\object{GRB 990705} & 56.2 & 0.46 & 82.5 & 0.44 &109.1 & 0.41 & 247.8 & 0.43 \\
\object{GRB 990908} & 46.4 & 0.72 & 42.4 & 0.77 & 24.4 & 0.86 & 113.2 & 0.77 \\
\object{GRB 000528} & 22.7 & 0.56 & 48.5 & 0.53 & 48.9 & 0.53 & 120.1 & 0.54 \\
\hline \multicolumn{9}{c} { }
\end{tabular}
\end{table*}

\subsubsection{\object{GRB~000528}}
This GRB was detected on May 28, 2000 at 08:46:35 UT by \sax,
Ulysses and NEAR, and localized to better than 2$'$ (Gandolfi et
al. 2000d). The LC in the 2--5 keV band appears dominated by the
SC, whereas narrower peaks become more evident at higher energies
(Fig.~\ref{fig17}, upper left panel). Note that also SC is growing
at 5--10 keV and more at 10-26 keV. After evaluating the SC for
the 2--5 keV LC and retrieving from it the SC profile for the
other energy ranges, we realized that to fit the 10--26 keV data
we had to anticipate the starting time by 10 s, according to
Eq.~(\ref{FC}).
Note from Table 2 that for this GRB the SC amplitudes at energies
higher than 5 keV are twice bigger than at 2--5 keV:  $
k_{21}=k_{31}=2$.
There is a hint that above 10 keV the SC can show a faster decay
or can be structured. Possibly it could be better described by the
sum of two functions, like in the case of \object{GRB~990123}.

\section{SC and FC spectral properties}

The above results show that in several cases we found that SCs
change with the energy differently than the total prompt emission signal
and therefore the spectral properties of SCs and FCs cannot be
similar to those of the original LCs. The simplest way to describe
these changes is to compare the hardness ratios of FCs and SCs
with the original ones. All the HR values are given in
Table~\ref{tab4} and in Table~\ref{tab5} are reported the
correspondent photon indices, derived using the method described
in Sect. 2.1. The values of HR$_2$ and $\Gamma_2$ for the GRBs
with  a soft SC given in square brackets were computed using the
5--10 keV FC and the original 10--26 keV LC. In the following we
will consider mainly HRs between the time integrated counts at
5--10 keV and 2--5 keV, because HR$_2$ values in the case of soft
SCs were not evaluated.

\begin{figure}
\centering
      \hspace{0cm}
\includegraphics[height=9.cm,angle=0]{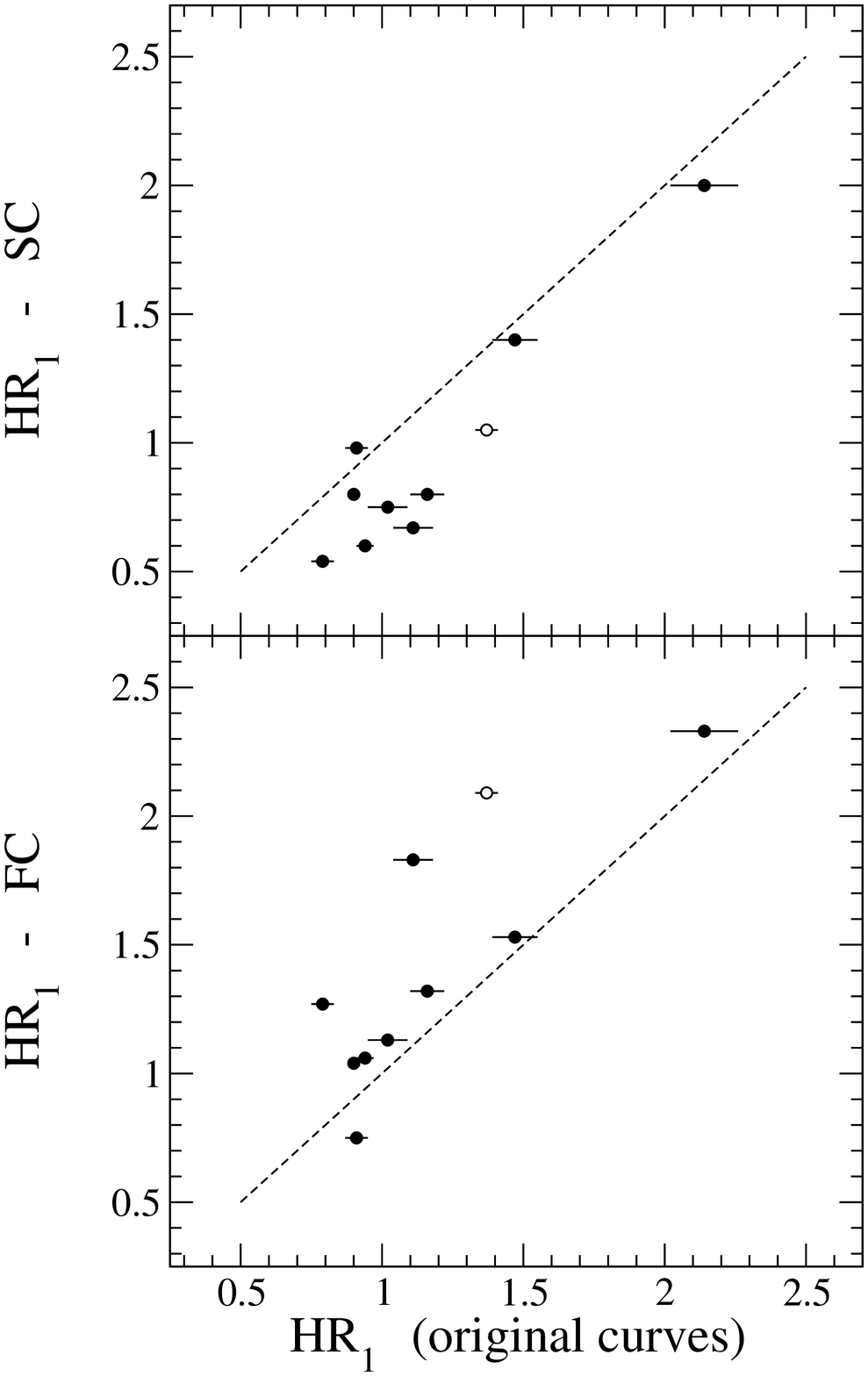}
\caption{ Hardness ratios of the SCs (upper panel) and FCs (lower
panel) plotted against those of the original light curves of the
eleven studied GRBs. Open circle indicates GRB~990123 which showed
the largest change.}
\label{fig18}
\end{figure}

Fig.~\ref{fig18} shows how HRs of the slow and fast components
compare with those of the original data sets: the dashed diagonal
line corresponds to unchanged HRs.
Note that, while SC points generally lie below the diagonal line,
those of FCs are on the opposite side. The unique exception is
\object{GRB~990908} that has an SC moderately harder than the original data set.
For the majority of GRBs spectral differences are not large, with 
the exception of \object{GRB 990123} and \object{GRB 001011}.
The mean value of HR$_1$ for the original LCs of GRBs with a soft SC
is 0.99, while that of SCs is 0.69 and the mean of differences is
0.25.
Spectral behaviour of FCs is symmetric with an increase of the mean
HR$_1$ to 1.27 and a mean difference of 0.27.

The corresponding variations for GRBs having hard SCs are different.
Note first that this subsample is not so homogeneous as the former:
HR values of OR data are generally higher with the only exception of
\object{GRB~990908} whose HRs are similar to those of bursts with soft SCs.
HR$_1$ values of SCs are only moderately smaller than OR data,
again with the exception of \object{GRB~990908}. We recall that this GRB
is somewhat peculiar because of the absence of prominent features
in its LC, in fact it has also the softest FC in all the sample.

Another peculiar behaviour is that of \object{GRB~000528}, which has the highest
values of HR$_1$ whereas those of HR$_2$ are comparable to other GRBs.
As already noticed in Sect. 4.2.5, LCs in Fig.~\ref{fig17} show that
it is dominated by the SC in all the energy bands, particularly
at 2--5 keV where the FC is slightly above the background level,
while some peaks become apparent in the other LCs.
We cannot exclude that high HR values can be related to the
difficulty of the SC modelling, however, a possible explanation to
be investigated in more detail is that of the large fraction of photons
below a few keV are absorbed in the source itself or in its environment.

\begin{table*}
\centering
\vspace{1cm}
  \caption{Hardness Ratios of slow and fast components and of original
            GRB light curves. Numbers in round brackets are 1$\sigma$
            statistical errors while numbers in square brackets are
            the HR$_2$ for GRBs without high energy SC measured between the
           FC at 5--10 keV and the OR at 10--26 keV.}
  \label{tab4}
  \begin{tabular}{c|cc|cc|cc}
    \hline
GRB & ~~~~~~~~SC  &   & ~~~~~~~~~~~~FC & & ~~~~~~~~~~~~~~~~~OR  &
\\ &HR$_1$ & HR$_2$ & HR$_1$ & HR$_2$ &  HR$_1$ & HR$_2$   \\
\hline
\object{GRB 980519} & 0.60 &    & 1.06 & [0.78] & 0.94 (0.03) & 0.63 (0.03)\\
\object{GRB 990704} & 0.54 &      & 1.27 & [0.91] & 0.79 (0.04) & 0.50 (0.03)\\
\object{GRB 001011} & 0.67 &      & 1.83 & [1.28] & 1.11 (0.07) & 0.81 (0.06)\\
\object{GRB 001109} & 0.75 &      & 1.13 & [0.95] & 1.02 (0.07) & 0.76 (0.07)\\
\object{GRB 010222} & 0.80 & 0.33 & 1.04 &  1.18  & 0.90 (0.02) & 0.71 (0.02)\\
\object{GRB 010412} & 0.80 &      & 1.32 & [1.13] & 1.16 (0.06) & 0.89 (0.05) \\ 
\hline
\object{GRB 990123} & 1.05 & 0.88 & 2.09 & 1.21 & 1.37 (0.04) & 1.04 (0.04) \\
\object{GRB 990705} & 1.40 & 1.21 & 1.53 & 1.41 & 1.47 (0.08) & 1.32 (0.06) \\
\object{GRB 990908} & 0.98 & 0.64 & 0.75 & 0.35 & 0.91 (0.04) & 0.58 (0.04) \\
\object{GRB 000528} & 2.0  & 1.0  & 2.33 & 1.3 & 2.14 (0.12) & 1.02 (0.05) \\
\hline
\end{tabular}
\end{table*}

\begin{table*}
\centering
\vspace{1cm}
  \caption{Photon spectral indices of the SC, FC and the original light
           curves. Numbers in round brackets are 1$\sigma$
            statistical errors while numbers in square brackets are
            the HR$_2$ for GRBs without high energy SC measured between the
           FC at 5--10 keV and the OR at 10--26 keV.}
  \label{tab5}
  \begin{tabular}{c|cc|cc|cc}
    \hline
GRB & ~~~~~~~~SC &  & ~~~~~~~~~~~~FC &  & ~~~~~~~~~~~~~~~~~OR & \\
 & $\Gamma_1$ & $\Gamma_2$ & $\Gamma_1$ & $\Gamma_2$ & $\Gamma_1$ & $\Gamma_2 $ \\
\hline
\object{GRB 980519} & 1.56 &  & 1.16 & [1.39] & 1.24 (0.02) & 1.51 (0.04) \\
\object{GRB 990704} & 1.42 &  & 0.81 & [1.05] & 1.15 (0.04) & 1.47 (0.05)\\
\object{GRB 001011} & 1.45 &  & 0.73 & [0.99] & 1.09 (0.05) & 1.32 (0.06) \\
\object{GRB 001109} & 1.21 &  & 0.93 & [1.05] & 0.99 (0.05) & 1.21 (0.07) \\
\object{GRB 010222} & 1.18 &1.80& 0.99 & 0.94 & 1.10 (0.01) & 1.26 (0.02) \\
\object{GRB 010412} & 1.15 &  & 0.80 & [0.91] & 0.89 (0.04) & 1.07 (0.04) \\
\hline
\object{GRB 990123} & 0.94 & 1.06 & 0.45 & 0.84 & 0.74 (0.02) & 0.94 (0.03) \\
\object{GRB 990705} & 1.01 & 1.12 & 0.95 & 1.01 & 0.98 (0.04) & 1.06 (0.04) \\
\object{GRB 990908} & 1.18 & 1.47 & 1.37 & 1.92 & 1.22 (0.03) & 1.55 (0.05) \\
\object{GRB 000528} & 0.65 & 1.13 & 0.54 & 1.12 & 0.60 (0.04) & 1.13 (0.04) \\
\hline
\end{tabular}
\end{table*}

The most interesting case is that of the famous \object{GRB~990123} whose
FC shows the highest increase of HR$_1$: we found a value of 2.09
while that of the OR data is only 1.37, the corresponding ratio
for the SC lowers only to 1.05. Note also that the changes of
HR$_2$ are much smaller.
The recent spectral analysis of Corsi et al. (2005) indicates
that the low-frequency extrapolation of the WFC spectrum of the
 prompt emission fails to match the simultaneous optical photometric data.
The best fit photon index given by these authors is 0.89$\pm$0.08,
intermediate between our $\Gamma_1$ and $\Gamma_2$ (see Table 5).
As written in Sect. 4.2.1 the extrapolation of the X-ray spectrum
of the prompt emission in the optical range is much lower than the fluxes
given by Akerlof et al. (1999a,b); one could then try to associate
this emission with one of the GRB components.
From the spectra given by Corsi et al. (2005), however, we can see
that an X-ray spectrum able to match optical points must have a
photon index close to 2. We found, instead, photon indices of both
components close to unity, too hard to avoid this discrepancy.

\begin{figure}
      \hspace{0cm}
\includegraphics[height=8.0cm,angle=-90]{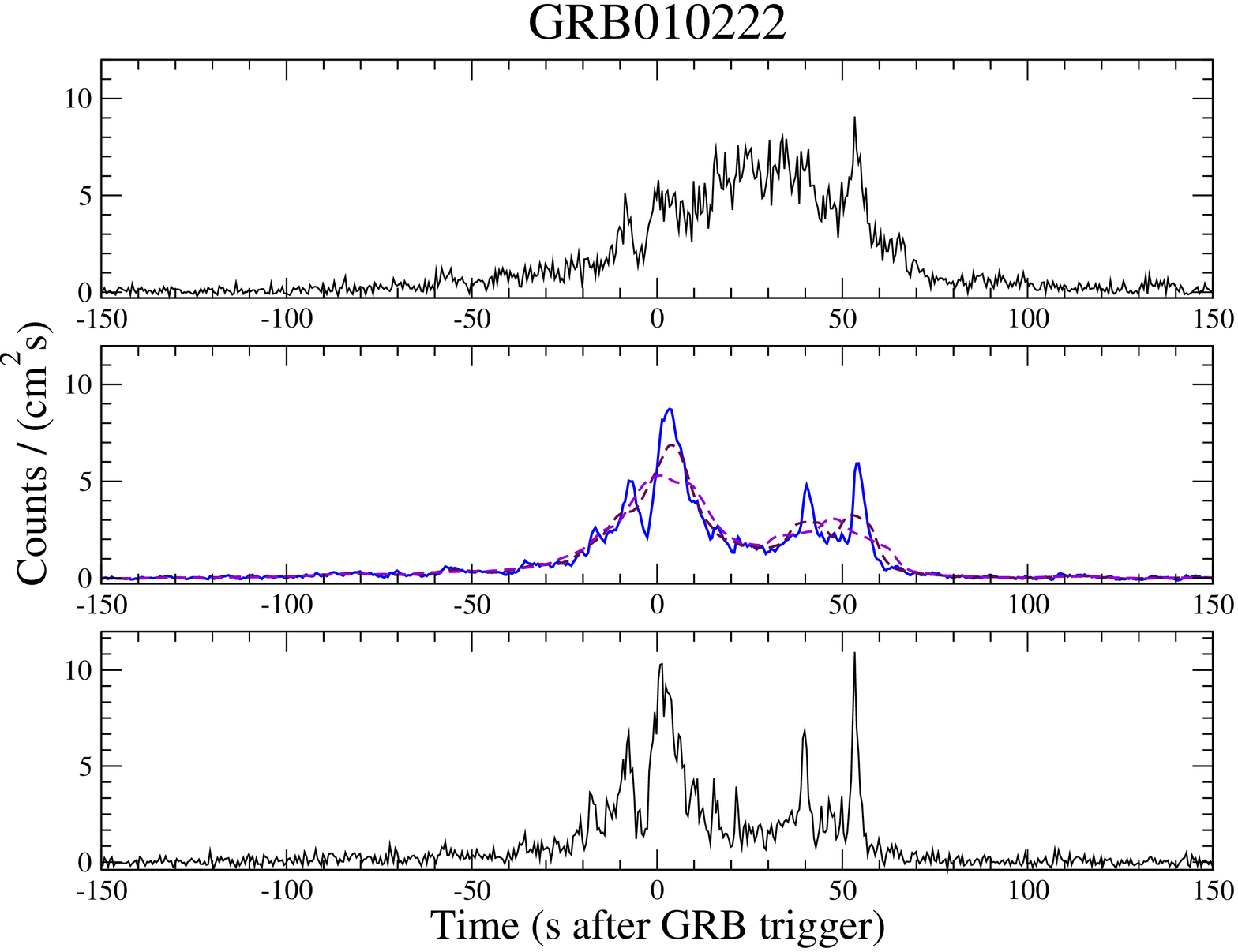}
\caption{Top panel: GRB~010222 light curves at 2--5 keV.
Middle panel: modified light curves of GRB~010222 at 2--5 keV
obtained smoothing the 10-26 keV light curve with the scale factor
derived from Eq.~(\ref{Fenimore}) (solid line) and considering
even larger values (dashed lines).
Bottom panel: original light curve of GRB~010222 at 10--26 keV. }
\label{fig19}
\end{figure}

\section{The nature of the slow component}

Our analysis of X-ray LCs of a sample of GRB detected by WFCs on
board {\it Beppo}SAX has shown that their complex shapes can be
accounted  by the presence of a slow component. Before speculating
about the origin of this component we have to verify if an SC, like
those modeled in previous sections, can be produced by peak
blending as a consequence of the low energy broadening mentioned
in Sect.3. A possible consequence of this low energy broadening is
that peaks with a small time separation can blend together and
produce an apparent underlying pedestal evolving on a longer time
scale. To evaluate the relevance of this effect we measured the
FWHM of the most intense peaks in the three energy bands and
derived the scale factor for the broadening according to
Eq.~(\ref{Fenimore}). We then smoothed the 10--26 keV LC
with this factor resulting in a curve that represents the
2--5 keV temporal evolution if it would depend on energy only as
Eq.~(\ref{Fenimore}). The resulting 2--5 keV LC for
\object{GRB~010222}, which has a high S/N ratio, is shown in the
middle panel of Fig. 19 (solid line): the comparison with the
original 2--5 keV LC (upper panel) did not show a SC of strength and shape
comparable to the original signal and therefore we concluded that
the peak broadening cannot be the origin of SCs. A further
possibility is that the appearance of a SC is related to a
broadening even larger than that expected from
Eq.~(\ref{Fenimore}). We therefore smoothed the previously obtained
curve with running average filters
and found the light curves shown in
the central panel of Fig. 19 (dashed lines). It is
evident that all these LCs have very different structures from
the original 2--5 keV LC depending on the spectral properties
of individual features. Note also that there is no way to obtain
a SC peaking around $t$ = 30 s.

Spectral differences between FCs and SCs suggest that their
emission can be due to different emission mechanisms: one producing
highly variable structures and another responsible of the SCs. We
tried to investigate whether this second mechanism could be the
initial phase of the afterglow but failed to find evidence that SC
spectra are similar to those of the afterglows. The latters are
usually characterised by photon indices $\Gamma$ $\simeq$ 2 (De
Pasquale et al. 2005) while the medium $\Gamma$ values of SCs are
all smaller (see Table~\ref{tab5}).
Furthermore, SCs are generally softer than the corresponding FCs
and their spectra are not single power laws but steepen at
high energy, as indicated by the different values of $\Gamma_1$
and $\Gamma_2$ (when measured), at variance with the afterglow
spectra.

Another possible interpretation is that SC is originated
in an outflow photosphere as in the model developed by
M\'esz\'aros \& Rees (2000), M\'esz\'aros et al. (2002),
Rees \& M\'esz\'aros (2005).
Usually the highly variable emission of GRBs is attributed
to internal shocks occurring at a certain distance from the
centre of the relativistic outflow, that is beyond the
photosphere at which the flow become optically thin.
At small enough radii the shock through dissipative effects can
create a number of pairs sufficient to reestablish a second
leptonic photosphere with a limiting radius beyond which the shock
remain optically thin. 
Thus above this radius there will be a region favorable for producing
the highly variable signals while the region below it will be favorable
to produce a less variable emission.
Moreover, sub photospheric dissipation leads to an increase of
the radiative efficiency of the outflow, consequently boosting
the quasi-thermal photospheric component so that it can dominate
the synchrotron component due to non-thermal shocks outside the
photosphere.
The expected spectrum of this thermal component is that of Black Body
(BB), but it can also show a Comptonized component (M\'esz\'aros and
Rees 2000). The $kT$ of the BB component depends on the dimensionless
entropy $\eta = L/\dot{M}c^2$ (here $L$ is the fireball luminosity and
$\dot{M}$ the mass rate of the outflow) and on the time scale variability
$\xi = t_v/t_0$ with $t_0 = 3.25\times10^{-3}\mu_1$ s the Kepler rotation
timescale ($\mu_1$ is the black hole mass in units of 10 solar masses).
In particular $kT$ values ranging from a few keV to tens of keV are obtained
for $\eta < 100$ and $\xi > 10^3$, the latter corresponding to long duration
GRBs, like those in our sample.

In this scenario it is natural to associate the SC with the photospheric
emission and FC with the more variable and harder non-thermal emission.
It is important, therefore, to verify that the SC is or not a trace of a
photospheric GRB emission, but this require the development of a
complex physical model that is beyond the aim of this work.
In any case some useful indications for this modelling can be obtained
by the evaluation of the energy dependence of the luminosity ratio
between the components.
As for example, we considered the two possible photon spectral
distributions (in the observers' frame) of these two components shown
in Fig. 20, where we plotted a BB spectrum with $kT$ = 1.2 keV and
a slowly steepening spectrum representing the flatter time averaged
synchrotron emission from the peaks of the FC, or, if present the
Comptonized component.

Using these simple spectra we computed the fractional content of the
SC in the same energy ranges considered in our analysis and found
0.63 (2--5 keV), 0.29 (5--10 keV) and 0.03 (10--26 keV).
These numbers compare well with those of some GRBs with a soft SC
in our sample, for instance they are very close to those measured for
\object{GRB~001011}.

In the case of \object{GRB~990704} we found that the duration of the SC at 5--10
keV is significantly shorter than at 2--5 keV and satisfy the
Fenimore et al. (1995) relation (Sect. 4.1.2).
It is unclear whether photospheric light curves do or do not have
the same time evolution at different energies, as pointed out in 
Sect. 4.1.2.. In the case it was the same, one has to invoke other 
emission processes and/or geometries to explain such a peculiar behaviour.

\begin{figure}
      \hspace{0cm}
\includegraphics[height=8.0cm,angle=-90]{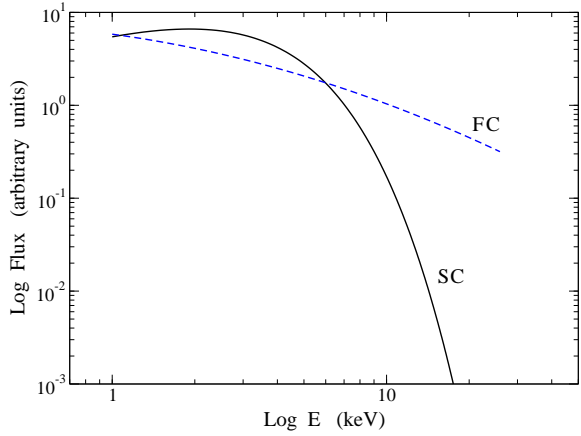}
\caption{Possible energy distributions of a SC with a soft spectrum
and a FC with a hard spectrum. The SC spectrum (solid line) is a
black body with $kT$ = 1.2 keV, while that of FC (dashed line)
is a flat non-thermal spectrum steepening at high energies.}
\label{fig20}
\end{figure}

\section{Conclusions}
Our analysis of X-ray LCs of a sample GRBs detected by WFCs on board
{\it Beppo}SAX provided evidence of components evolving on time scales
of several tens/hundreds of seconds.
Although in some cases the shape of the SC cannot be well defined
because of the low S/N ratio, for the majority of GRBs in our sample,
the existence of a SC can be considered safely established.
In particular, in the case of \object{GRB~010222} we verified that SCs are not 
originated by blending of narrow peaks and of features closely spaced 
in time as a consequence of the low-energy broadening found by Fenimore 
et al. (1995).
Likely the same conclusion holds for the other GRBs.
Furthermore, the subtraction of a SC model from original LCs gave
residual fast components showing time structures very similar to
those of LCs above 10 keV and having peak broadening in agreement
with Eq.~(\ref{Fenimore}).

Spectral analysis on SCs was not simple and we were able to derive
some indicative photon indices. In six of GRBs studied by us, the
spectra of SCs were characterised by a rather sharp cut-off above
$\sim$10 keV.
In a few cases, however, we found a clear indication
that spectra of SCs are harder with a detectable emission in the
range 10--26 keV.
A very promising scenario to explain these spectral behaviours seems
that of a thermal component originating from an expanding photosphere
having $kT$ of a few keV as in the model proposed by M\'esz\'aros
\& Rees (2000).
To verify this hypothesis it is necessary to develop a more detailed
spectral model to be adapted to individual GRBs.
This is beyond the present work, which is essentially based on the
study of LCs, and it can be a further step for the understanding of
such complex transient phenomena.

It is also relevant to develop a fast algorithm to be applied for
a safe identification and modelling of slow components in GRB
light curves. The simple approach adopted by us in this work,
although useful in a first level analysis, is user dependent
and therefore cannot be of general application. A filtering
technique, based on wavelet analysis, considering the power
distribution over the various time scales can be useful, but its
development and testing over a large variety of LC profiles,
is not a simple task.

As a final remark we stress the relevance of broad band GRB
observations. The comparison between light curves at different
energies is an unavoidable step to unravel the presence of
components evolving on different time scales. It is important
that future space missions mainly devoted to GRB observations
will be designed to cover also the X-ray bands down to a fraction of
keV.

\begin{acknowledgements} We are grateful to the referee J.T. 
Bonnell for his comments and suggestions.
This research was supported by INAF and Universit\'a La Sapienza di Roma.
The authors are grateful to ASDC, ASI Data Center, in particular to
F. Verrecchia for making the WFC data available and the support in the 
data transfer, and to  L. Amati, A.Corsi and M. Feroci for providing us 
some data.

\end{acknowledgements}

\newpage


\begin{thebibliography}{ }
\bibitem[Akerlof et al. 1999a]{aker99a} Akerlof, C.W., et al. 1999a, GCN
Circ. 205 
\bibitem[Akerlof et al. 1999b]{aker99b} Akerlof, C.W., et al. 1999b,
Nature, 398, 400
\bibitem[Akerlof \& McKay 1999]{aker&mc}Akerlof, C.W., McKay, T. A.,1999,
IAU Circ. 7100
\bibitem[Amati et al. 2000a]{amati00a}Amati, L., Frontera, F., et
al. 2000a, IAU Circ. 7519
\bibitem[Amati et al. 2000b]{amati00b}Amati, L., Frontera, F., et 
al. 2000b, Science, 290,953
\bibitem[Amati et al. 2002]{amati02}Amati, L. et al., 2002, A\&A 390,
81
\bibitem[Arnaud et al. 1996]{arnaud}Arnaud, K.A., 1996, Astronomical Data
Analysis Software and Systems V, eds. Jacoby G. and Barnes J.,
p17, ASP Conf.
 Series volume 101
\bibitem[Berger \& Frail 2001a]{berger01a} Berger, E., Frail, D.A. 2001a,
GCN Circ. 1168
\bibitem[Berger \& Frail 2001b]{berger01b} Berger, E., Frail, D. A. 2001b,
GCN Circ. 968 
\bibitem[Boella et al. 1997]{boella97} Boella, G., Butler, R.C. et al., 1997,
A\&AS  125, 557
\bibitem[Castro Cer\'on et al. 2004]{castro04} Castro Cer\'on, J.M., 
Gorosabel, J.,  et al., 2004, A\&A , 424, 833
\bibitem[Celidonio et al. 1999]{celid99}
Celidonio, G., Tarei, G., et al. 1999,
IAU Circ. 7218
\bibitem[Corsi et al. 2005]{corsi05} Corsi A., Piro L., et al.,
2005, A\&A , in press, astro-ph/0504607
\bibitem[De Pasquale et al. 2005]{depasq05} De Pasquale, M., Piro, L., 
Gendre, B.  2005, A\&A ,submitted
\bibitem[Di Paola et al. 2001]{dipaola01} Di Paola,~A., Antonelli,~L.A.
et al. 2001,~GCN~ Circ. No.~977
\bibitem[Djorgovski et al. 1998]{Djor98} Djorgovski, S.G., Gal, R. R.,
Kulkarni, S. R., Bloom, J. S., Kelly, A. 1998, GCN Circ. 79
\bibitem[Fenimore et al. 1995]{fenimore95} Fenimore, E.E., in 't Zand, J.J.,
 Norris, J. P., Bonnell, J. T., Nemiroff, R. J. 1995, ApJ, 448, L101
\bibitem[Feroci et al. 1998]{feroci98}Feroci, M., Piro, L., Daniele,
M.R. et al., 1998, IAU Circ. 6909
\bibitem[Feroci et al. 1999]{feroci99} Feroci, M., et al. 1999, IAU Circ. 7095
\bibitem[Feroci et al. 2001]{feroci01}Feroci, M., Antonelli, L. A.,
Soffitta, P. et al. 2001,  A\&A 378, 441F
\bibitem[Fich et al. 2001]{fich01} Fich,~M., Phillips, R.R. et al., ~2001,
GCN~Circ. 971
\bibitem[Fishman et al. 1985]{fish85} Fishman, G., et al. 1985, ICRC, 3, 343F
\bibitem[Fishman et al. 1992]{fish92} Fishman, G., et al. 1992, in Gamma-Ray
 Bursts: Huntsville, 1991, ed. W. S. Paciesas \& G. J. Fishman (New York: AIP),
13
\bibitem[Frail et al. 1999]{frail99}Frail, D. A., et al. 1999, GCN Circ. 211
\bibitem[Frontera et al. 1997]{front97} Frontera, F., Costa, E. et al., 1997,
A\&AS  122, 357
\bibitem[Frontera 2004]{front04} Frontera, F., 2004, ASP Conference Series, 
Vol. 312
\bibitem[Fruchter et al. 1999]{fruch99} Fruchter, A., et al. 1999, ApJ,
519, L13
\bibitem[Galama et al. 1999a]{Galama99a} Galama, T., et al., 1999a, Nature,
398, 394
\bibitem[Galama et al. 1999b]{Galama99b} Galama, T., et al., 1999b, GCN Circ.
212 
\bibitem[Gandolfi et al. 2000a]{Gand00a} Gandolfi,~G.~et al., 2000a,~
GCN~Circular No 846
\bibitem[Gandolfi et al. 2000b]{Gand00b} Gandolfi,~G., Piro,~L. 2000b,
GCN~Circular No. 878
\bibitem[Gandolfi et al. 2000c]{Gand00c} Gandolfi, G., Piro, L. 2000c,
GCN~Circular No. 879
\bibitem[Gandolfi et al. 2000d]{Gand00d} Gandolfi,~G., 2000d, ~BeppoSAX Mail
n.~00/10~and~00/11 (Rome:~IASF)
\bibitem[Gorosabel et al. 2000]{Goro00} Gorosabel, J, Hjorth, J., Pedersen, H.
 et al. 2000, GCN Circular No. 849
\bibitem[Gorosabel et al. 2002]{Goro02} Gorosabel, J, Fymbo, J.U., Hjorth, J.,
 et al.. 2002, A\&AS 384, 11
\bibitem[Heise et al. 1999a]{Heise99a} Heise, J., in 't Zand,
J., Celidonio, G., et al. 1999a, IAU Circ., No. 7217
\bibitem[Heise et al. 1999b]{Heise99b} Heise, J., Delibero, C., et al. 1999b,
IAU Circ., No. 7099
\bibitem[Henden et al. 2001a]{Henden01a} Henden,~A. et al., 2001a,~GCN
Circ. No.~961
\bibitem[Henden et al. 2001b]{Henden01b} Henden,~A. et al.,~2001b, GCN~Circ.~96
\bibitem[Hjorth  et al. 1999a]{Hjorth99a} Hjorth, J., Pedersen, H.,
Jaunsen, A.O., Andersen, M.I. 1999a, A\&AS 138, 461
\bibitem[Hjorth  et al. 1999b]{Hjorth99b} Hjorth, J., et al., 1999b,
GCN Circ. 219
\bibitem[Holland  et al. 2000]{Holl00} Holland et al., 2000, GCN
Circ. 753
\bibitem[Jager et al. 1997]{jager97} Jager, R., Mels, W.A. et al., 1997,
A\&AS  122, 299
\bibitem[Jaunsen et al. 1998]{jau98} Jaunsen, A. O., Hjorth, J.
et al., 1998, GCN Circ. 78
\bibitem[Kelson et al. 1999]{kelson99}Kelson, D.D., Illingworth, G.D.,1999, IAU
Circ. 7096
\bibitem[Kulkarni et al. 1999a]{kulk99a} Kulkarni, S.R., Djorgovski,
S.G.,  et al. 1999a, Nature,398, 389
\bibitem[Kulkarni et al. 1999b]{kulk99b}Kulkarni, S.R., Frail, D.A., 1999b, ApJ
522, L97
\bibitem[Lachowicz \& Czerny 2005]{lacz05} Lachowicz P., \& Czerny B.,
2005, MNRAS (in press), astro-ph/0412136
\bibitem[Link, Epstein, \& Priedhorsky 1993]{link} Link, B., Epstein, R. I., \&
Priedhorsky, W. C. 1993, ApJ, 408, L81
\bibitem[Masetti et al. 2000]{Masetti} Masetti, N. et al. 2000, A\&A
354, 4730
\bibitem[M\'esz\'aros et al. 1993]{mesz93}M\'esz\'aros, P.,
Laguna, P., Rees, M.J., 1993, ApJ 415, 181
\bibitem[M\'esz\'aros \& Rees 2000]{mesz00}M\'esz\'aros,\& P., Rees,
 M.J., 2000, ApJ 530, 292
\bibitem[M\'esz\'aros et al. 2002]{mesz02}M\'esz\'aros, P., Ramirez-Ruiz, E.
et al. 2002, ApJ 578, 812
\bibitem[Muller et al. 1998]{Muller98} Muller, J. M., et al. 1998, IAU
Circ. 6910
\bibitem[Nicastro et al 1998]{nica98} Nicastro, L. et al. 1998, IAU
Circ. 6912
\bibitem[Norris et al. 1996]{norris96}Norris, J.P., Nemiroff, R.J.
et al. 1996, ApJ, 459, 393
\bibitem[Norris et al. 2000]{norris}Norris, J.P., Marani, G.F. \& Bonnell,
J.T. 2000, ApJ, 534, 248
\bibitem[Odewahn et al 1999]{ode99} Odewahn, S.C., et al. 1999, GCN Circ. 201
\bibitem[Piro et al 1998]{piro98} Piro,~L.,~1998,~~GCN~Circ. No.~75
\bibitem[Piro et al 1999a]{piro99a} Piro, L., et al. 1999a, GCN Circ. 199
\bibitem[Piro et al 1999b]{piro99b} Piro, L., et al. 1999b, GCN Circ. 202
\bibitem[Piro et al 1999c]{piro99c} Piro, L., et al. 1999c, GCN Circ. 203
\bibitem[Piro et al 1999d]{piro99d} Piro, L., et al. 1999d, GCN Circ. 406
\bibitem[Piro  2001a]{piro01a}Piro, L. 2001a, GCN Circ. 959
\bibitem[Piro  2001b]{piro01b}Piro, L. 2001b, GCN Circ. 1033
\bibitem[Rees \& Meszaros 2005]{rees05}Rees, M.J., \& Meszaros, P., 2005, ApJ
in press, astro-ph/0412702
\bibitem[Reichart et al. 2001]{reich01}Reichart, D.E., Lamb, D.Q. et al. 2001,
ApJ, 552, 57
\bibitem[Sari \& Piran 1999]{sari99} Sari, R., \& Piran, T. 1999, ApJ,
517, L109
\bibitem[Shirasaki et al. 2003]{Shira} Shirasaki, Y., Kawai, N., et al.
2003, Proceedings op the SPIE, 4851, 1310
\bibitem[Stanek et al 2001a]{Stanek01a} Stanek, K.Z., Jha, S., McDowell, et al.
2001a, GCN Circ. 970 
\bibitem[Stanek et al 2001b]{Stanek01b} Stanek, K.Z., Garvanich, P.,
Jha, S., Pahre, M. 2001b, IAU Circ., No 7586
\bibitem[Taylor et al 2000]{taylor00} Taylor, G.B., Frail, D.A.,
Kulkarni, S.R. 2000, GCN Circ. 880
\bibitem[Tavani 1996]{tavani96} Tavani, M., 1996, ApJ 466, 768
\bibitem[Torrence and Compo 1998]{torco98} Torrence, C., Compo, G.P.
1998, BAMS 79 61
\bibitem[Vanderspek et al. 2004]{vader04} Vanderspek R.,
Sakamoto T., Barraud C., 2004, ApJ 617, 1251
\bibitem[Vestrand et al. 2005]{vest} Vestrand, W.T., Wozniak, P.R., 
Wren, J.A. et al. 2005, Nature, 435, 178
\bibitem[in 't Zand et al. 1999]{Zand99} in 't Zand, J.J.M.,
 Heise, J., van Paradijs, J., Fenimore, E.E. 1999, ApJ 516, L57

\end{thebibliography}
\end{document}